\documentclass[useAMS,usenatbib,usegraphicx]{mn2e}
\usepackage{amsmath}
\usepackage{amsfonts}
\usepackage{amssymb}
\usepackage{multirow}

\title[Time evolution of probability density function of gamma ray burst (GRB)] 
{Time evolution of probability density function of gamma ray burst (GRB) - 
a possible indication of turbulence origin of GRB}

\author[Nilay Bhatt, Subir Bhattacharyya]
{Nilay Bhatt$^{1}$\thanks{E-mail:nilayb@barc.gov.in}   
 Subir Bhattacharyya$^{1}$ \thanks{E-mail:subirb@barc.gov.in }
  \\
$^{1}$Astrophysical Sciences Division, Bhabha Atomic Research Centre, Mumbai 400085, India \\
}

\begin{document}


\pagerange{\pageref{firstpage}--\pageref{lastpage}} \pubyear{2008}

\maketitle

\label{firstpage}

\begin{abstract}
Gamma ray burst (GRB) time series is a non-stationary time series with all its statistical
properties varying with time. Considering that each GRB is a different manifestation of the 
same stochastic process we studied the time dependent as well as time averaged probability 
density function (\emph{pdf}) characterizing the underlying stochastic process. The \emph{pdf}s  
are fitted with Gaussian distribution function and it has been
argued that the Gaussian \emph{pdf}s possibly indicate the turbulence origin of GRB. The spectral and
temporal evolution of GRBs are also studied through the evolution of spectral forms, color-color 
diagrams and hysteresis loops. The results do not contradict the turbulence interpretation of GRB.
\end{abstract}


\section{Introduction}
Even after the three decades of extensive observations of gamma-ray burst (GRB) prompt emission
there is no conclusive idea about the energy dissipation mechanism and the radiative process responsible
for high energy emission (\cite{Ly09, Gh10}). In the \emph{standard model} of GRB, known as fireball 
model (\cite{Gd86, Pa86, SP90, MR93, RM92, SP95, KPS99, MR00, MRR02}), the bulk 
energy associated with the relativistic expansion of the fireball, is dissipated into the random 
energy of particles through internal shocks. Internal shocks are produced due to the time dependent
speed of the relativistic outflow (\cite{KPS97, RM94, SP97}). But it is now established that the internal shock model suffers
from different shortcomings, particularly, the low radiation efficiency, problem of baryon loading
and the large dissipation radius (\cite{NK09, LB10}). 

\noindent As possible alternatives of internal shock model, \cite{LB03} and \cite{Ly06}
proposed electromagnetic model of gamma-ray burst whereas \cite{NK09} discussed relativistic 
turbulence model for GRB. In electromagnetic model the GRB outflow is considered as Poynting flux
dominated wind (\cite{Ly06}). The energy dissipation and particle acceleraton occur through 
magnetic dissipation process,
mainly through magnetic reconnection. In relativitic turbulence model by \cite{NK09}, it is 
considered that the fluid in GRB out flow is relativistically turbulent. The rapid variability 
in the GRB lightcurve arise due to the relativistic random velocity fluctuations of radiation emitting
eddies formed in the turbulence. The model was successfully applied to explain the X-ray--gamma-ray and
optical emission during the prompt phase of the brightest \emph{Swift} gamma-ray burst GRB080319B
(\cite{KN09}).

\noindent A photospheric emission model is proposed by \cite{LB10} very recently. Generally 
the photosphere, associated with the fireball, is dominated by thermal radiation. It is assumed that the
electrons, energised by some shock or due to magnetic reconnection, Comptonize the thermal photons 
in an optically thick medium. The multiple Comptonization leads to a power-law photon distribution in the 
keV--MeV region whereas a thermal peak remains in the keV energy range. The main advantage of of this
model is that it does not require any efficient dissipation mechanism and the radiative efficiency of
this model is also very high. But there is no strong observational evidence of any thermal component in GRB
spectra.     

\noindent As far as the spectral studies are concerned, the majority of observed time averaged prompt
emission spectra of GRBs are fitted with Band function (\cite{Band93}) which, indeed, is no more 
than a mathematical 
function having no physical interpretation yet. It is generally believed that the high energy emission 
in the prompt phase is due to the synchrotron process (\cite{Ta96a, Ta96b}). But the detail 
study of BATSE GRBs shows that for many GRBs the low energy spectral slopes are not consistent 
with the theory of synchrotron emission (\cite{Pre00}).

\noindent Therefore to have deeper understanding on the physical processes occuring in gamma-ray bursts
it would be benificial to study the time evolution of GRB spectra. It is also important to look into
the GRB time series from a different perspective and characterise the underlying process. 
\cite{LP02, GCG02, Md06}
and several other authors studied the 
time evolution of GRB spectra for BATSE GRBs. \cite{LP02} studied a 128 ms spectra for a
large sample of GRBs observed by BATSE. The spectra were fitted with synchrotron emission model where the
spectra were calculated in three emission regime, such as, for isotropic pitch angle emission of electrons,
for small angle pitch angle distribution of electrons and for self-absorbed systems. Eventhough the 
spectral fitting were done and the time evolution of spectral index and normalization were studied, but
the physical description was not self-consistently developed within the framework of fireball model.
\cite{GCG02} also studied BATSE GRBs and fitted the time resolved the spectra with Band
function, broken power-law, thermal Comptonization and synchrotron shock model. It was concluded that
the behaviour of the time resolved spectra is not consistent with the time integrated spectrum. The low
exergy spectral indices also violates the synchrotron spectral limit.

\noindent In this paper we studied the probability density function (\emph{pdf}) and 
spectral evolution 
of four bright GRBs detected by \emph{Swift} and tried to find out if there is any systematics among those 
GRBs. In \cite{BSS00}, the authors mentioned that the individual burst is a random realization
of the same standard stochastic process. In any stochastic process there is an underlying \emph{pdf} which 
characterises the process. Therefore it is important to find out the \emph{pdf} from GRB time series to 
identify if it has any general characteristic. Since GRBs are highly non-stationary process, 
we calculated time-dependent as well as time averaged \emph{pdf} for selected GRBs. 
We also studied the spectral evolution, colour-colour and hardness-count rate correlations for those GRBs.
 
\noindent In Section 2 we give a brief description of the selected GRBs, in Section 3 we describe the 
analysis procedure of \emph{Swift}-BAT data. Results are described and discussed in Sections 4 and  
we conclude the paper in Section 5.

\begin{figure}
\begin{center}
\includegraphics[width=0.17\textheight, height=0.45\textwidth, angle=-90]{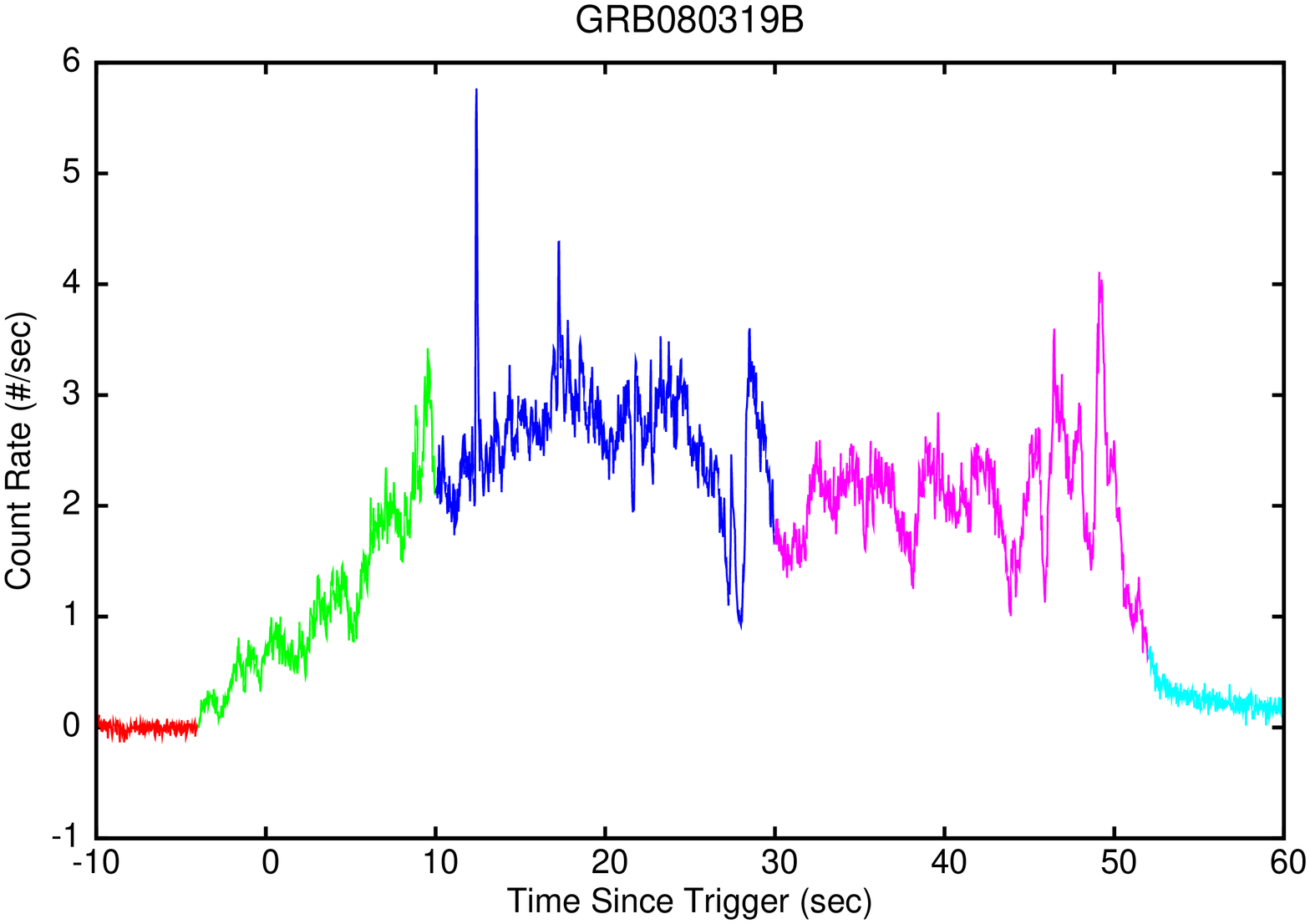}
\includegraphics[width=0.17\textheight, height=0.45\textwidth, angle=-90]{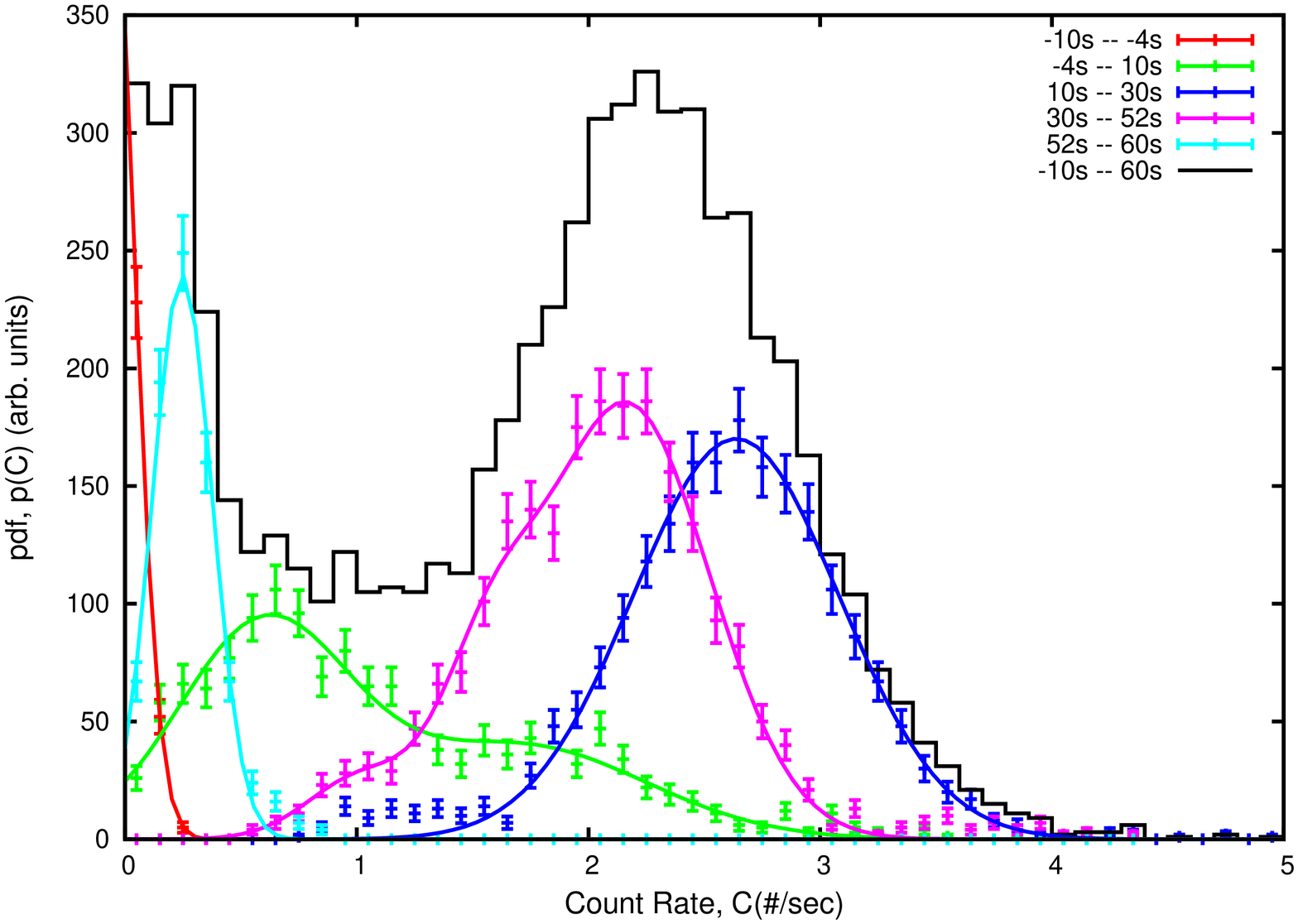}
\includegraphics[width=0.17\textheight, height=0.45\textwidth, angle=-90]{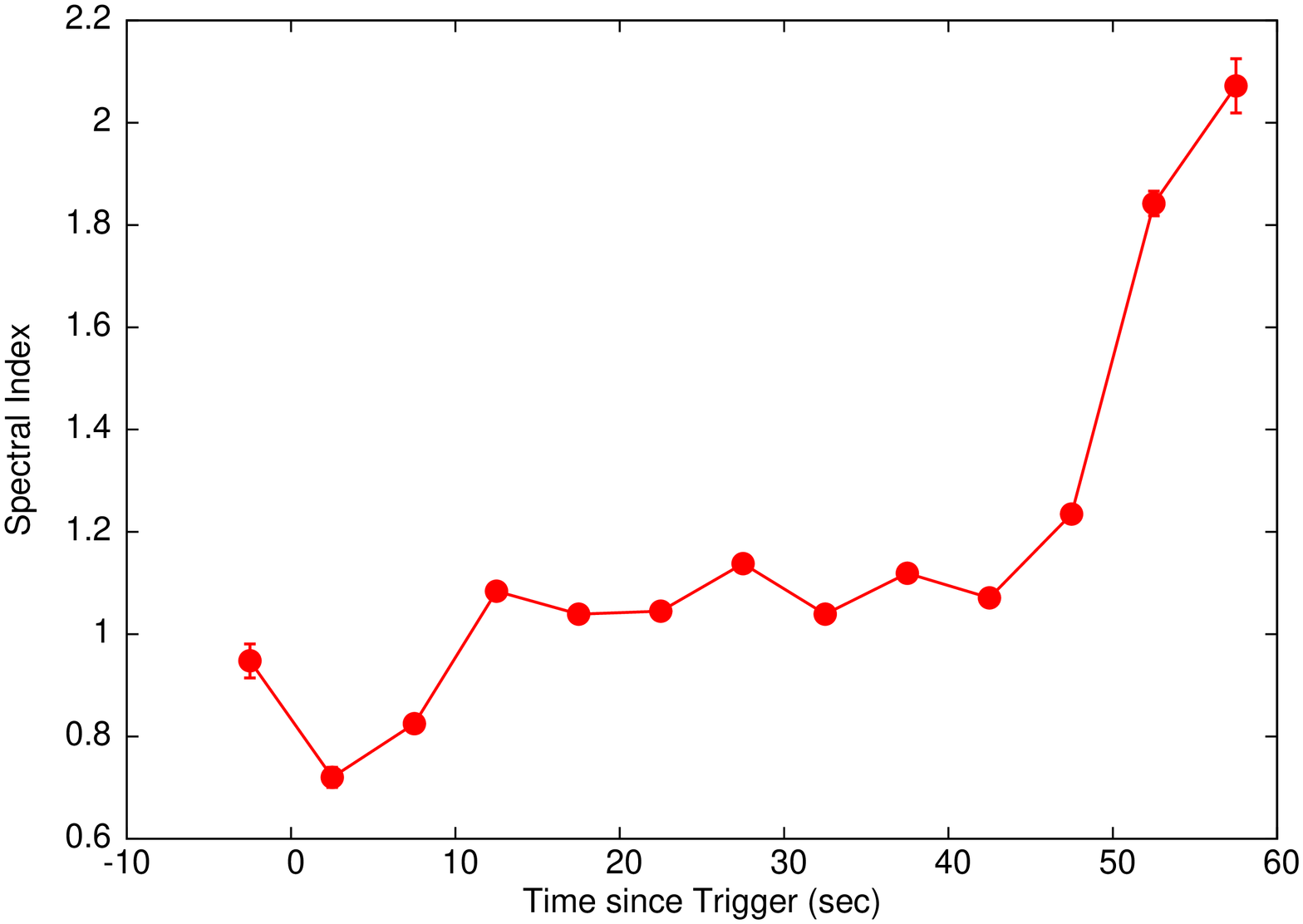}
\includegraphics[width=0.17\textheight, height=0.45\textwidth, angle=-90]{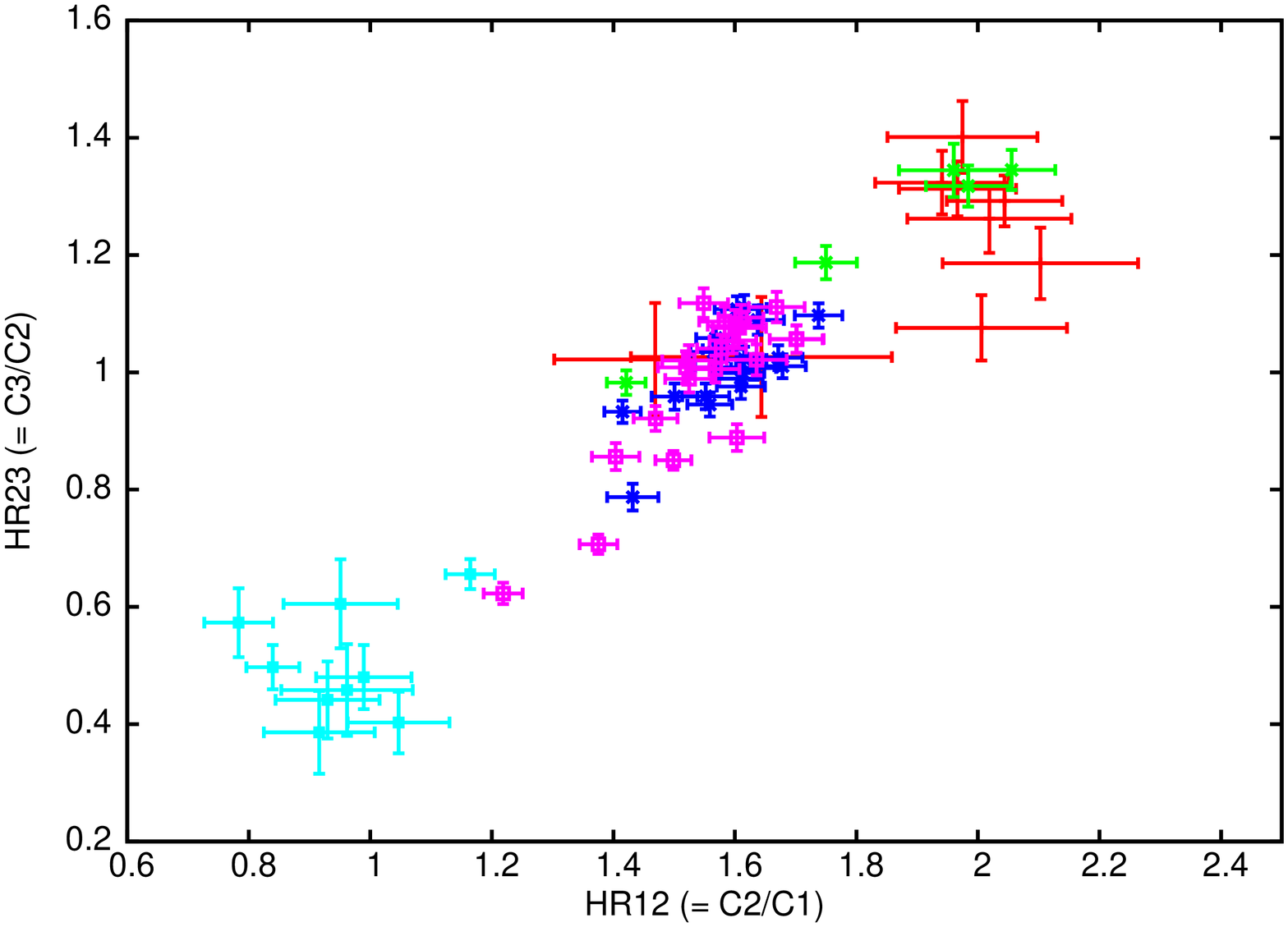}
\caption{\emph{First panel} (from top) : Lightcurve of GRB080319b; \emph{second panel} : probability density function
 - time averaged (black), time dependent (colours of different distributions correspond to the portions of the lightcurve with same colour);
 \emph{third panel} : variation of spectral index with time; \emph{fourth panel} : colour-colour diagram for the active
portion of the lightcurve (same colour convention maintained).} \label{pds1}
\end{center}
\end{figure}
\section{Selected GRBs}
We have analysed the \emph{Swift}-BAT data for GRB050525, GRB061121, GRB080319b and GRB080411. 
GRBs are selected based on their fluence. All four GRBs have fluence $> 10^{-5}$ erg cm$^{-2}$. 
For the sake of completeness we give a brief description each GRB below.
\begin{itemize}  
\item {\bf GRB050525 :} This GRB was observed with {\emph Swift}-BAT in the energy range 15--350 keV. It
ia at a redshift of 0.606. The observed fluence in the {\emph Swift}-BAT energy range is 
$1.86 \times 10^{-5}$ ergs cm$^{2}$. The time averaged spectrum in the 15--350 keV range was fitted
with a power-law ($\sim \nu^{-\alpha}$) with spectral index 0.83. The estimated isotropic energy is
$E_{iso}=1.94\times10^{52}$ erg (\cite{GCN050525a, GCN050525b}). 
\begin{figure}
\begin{center}
\includegraphics[width=0.17\textheight, height=0.45\textwidth, angle=-90]{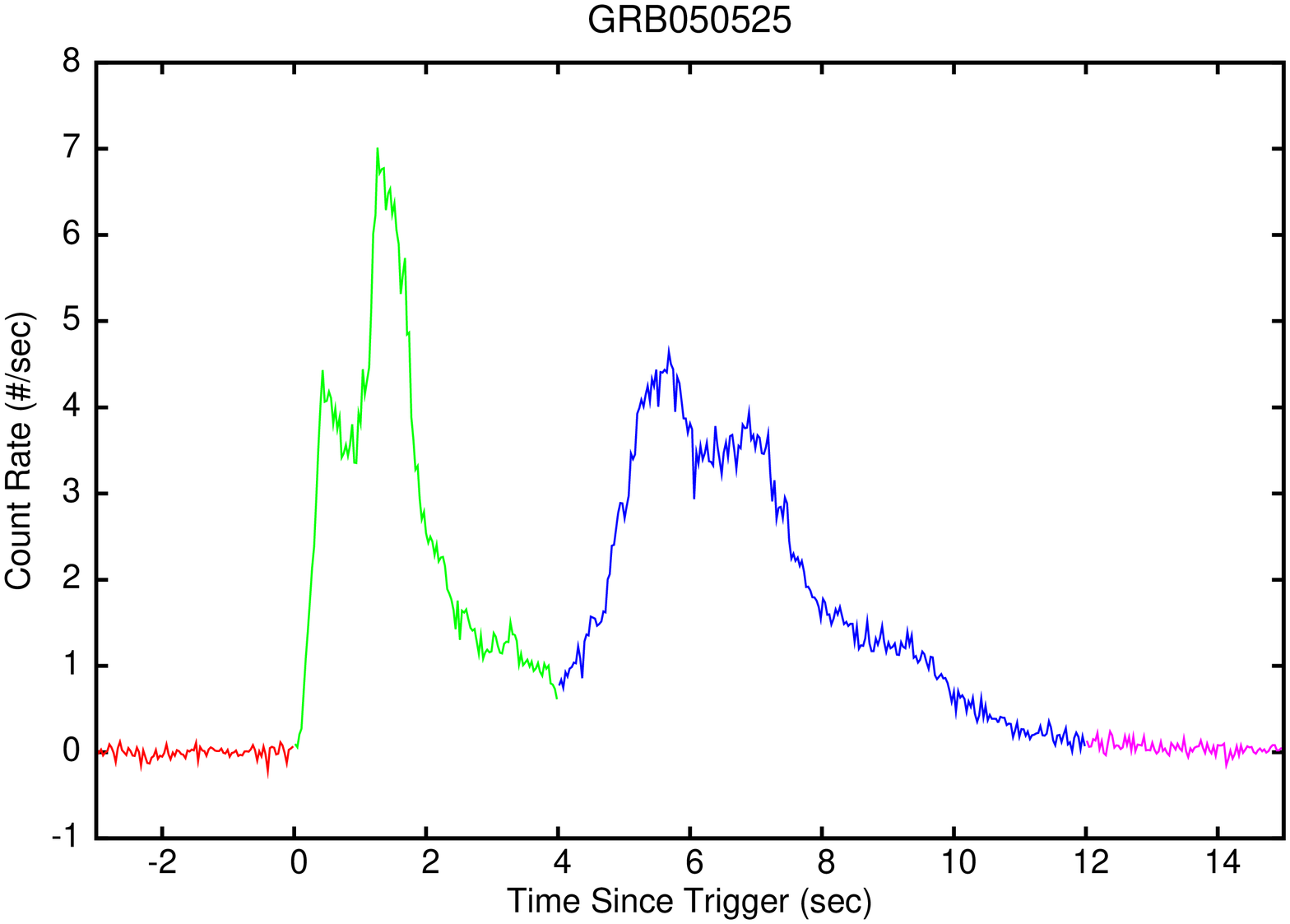}
\includegraphics[width=0.17\textheight, height=0.45\textwidth, angle=-90]{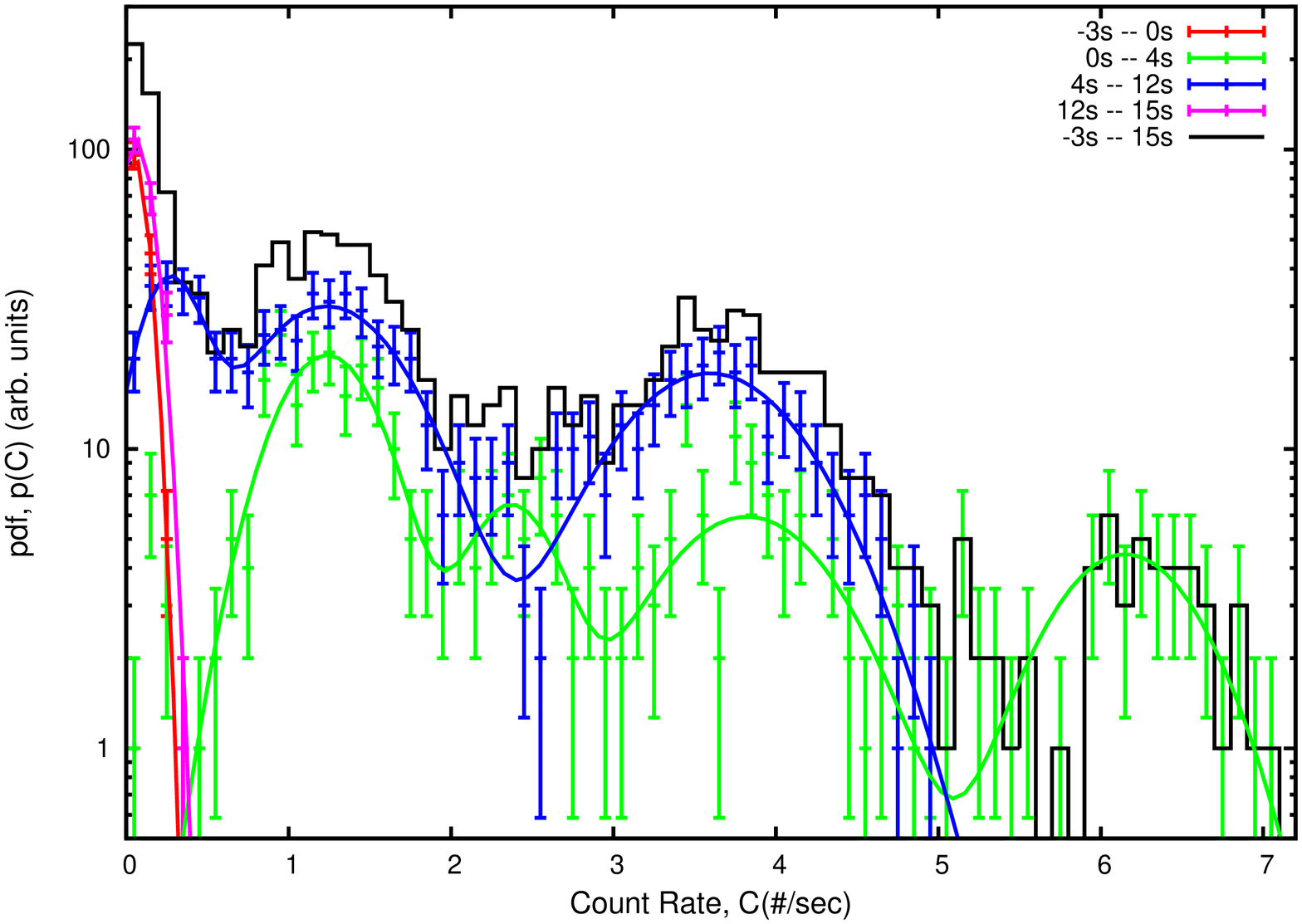}
\includegraphics[width=0.17\textheight, height=0.45\textwidth, angle=-90]{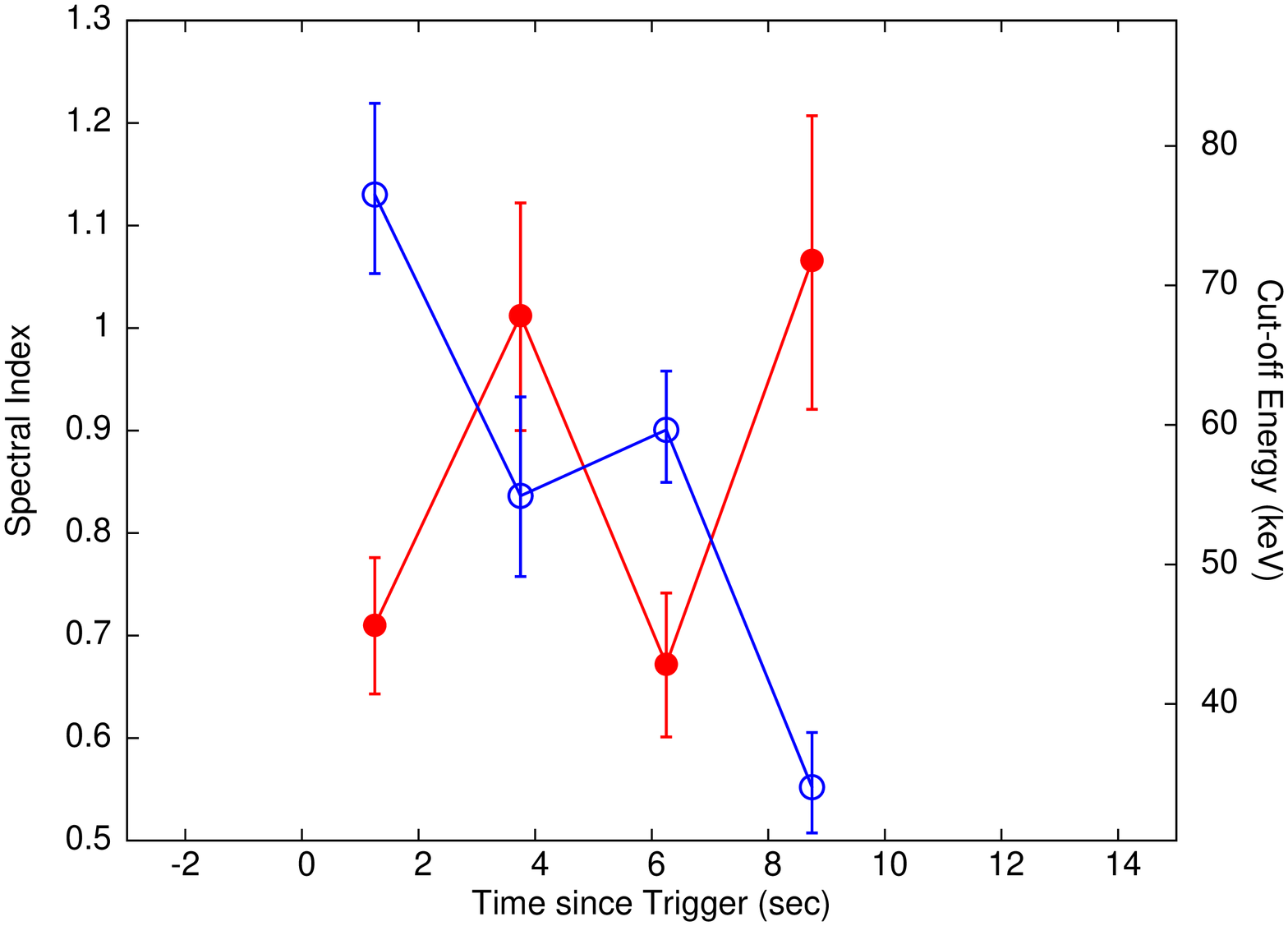}
\includegraphics[width=0.17\textheight, height=0.45\textwidth, angle=-90]{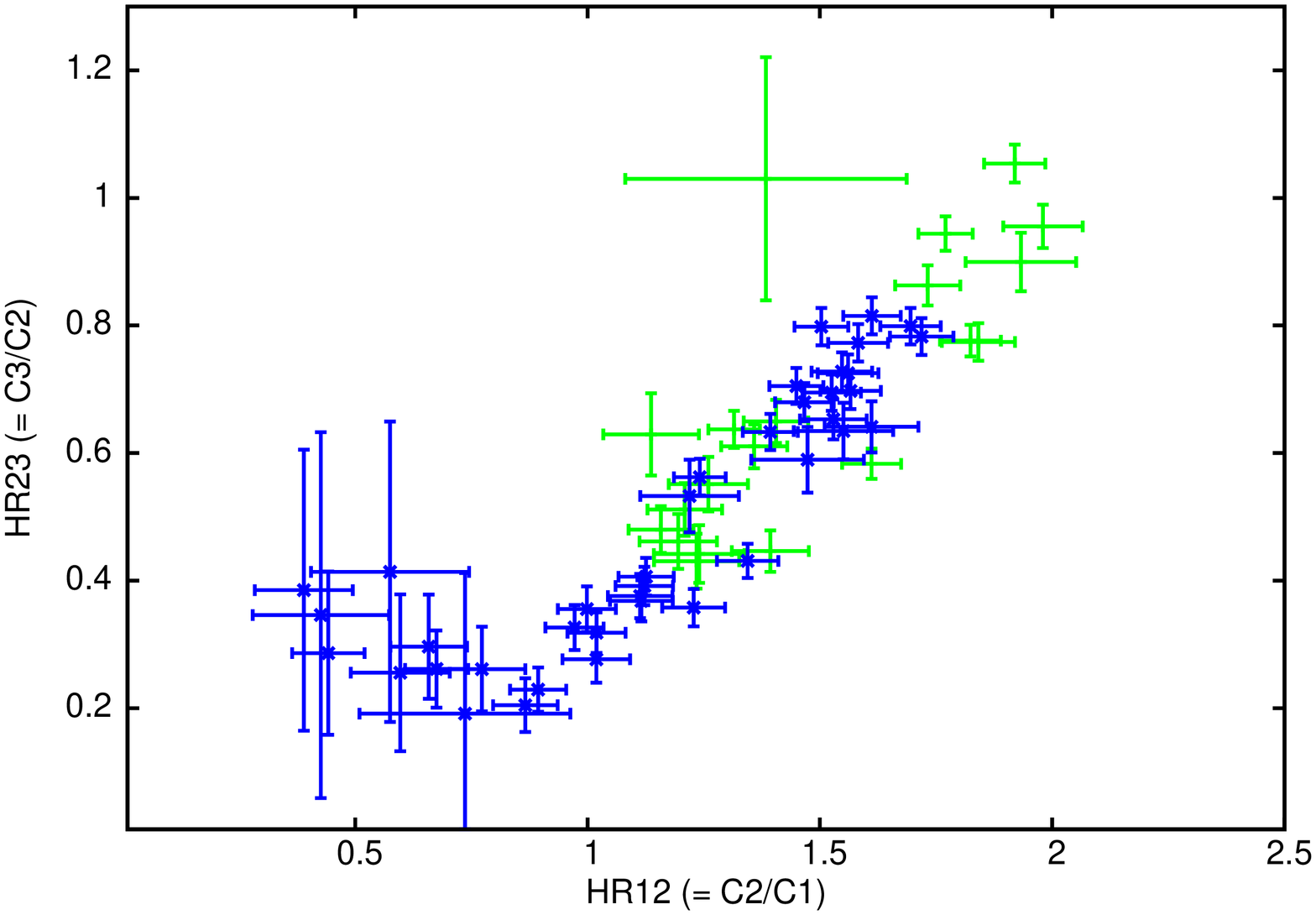}
\caption{Same as Figure \ref{pds1}, but for GRB050525.} \label{pds2}
\end{center}
\end{figure}
\item {\bf GRB061121 :} {\emph Swift}-BAT detected this burst alongwith a prescursor. Here we 
concentrate on the main event which has $T_{90}=18.2 \pm 1.1$ s. The {\emph Swift}/BAT spectrum 
was fitted with a power-law with spectral index 1.375. The measured fluance is $2.792\times10^{-5}$
erg cm$^{-2}$. The measured redshift for the burst is 1.314 which leads to a measured isotropic
energy $E_{iso}\approx6.7\times10^{53}$ erg. \cite{Page07} studied the broadband spectral
evoluton of the prompt and afterglow emission of the burst.   
\begin{figure}
\begin{center}
\includegraphics[width=0.17\textheight, height=0.45\textwidth, angle=-90]{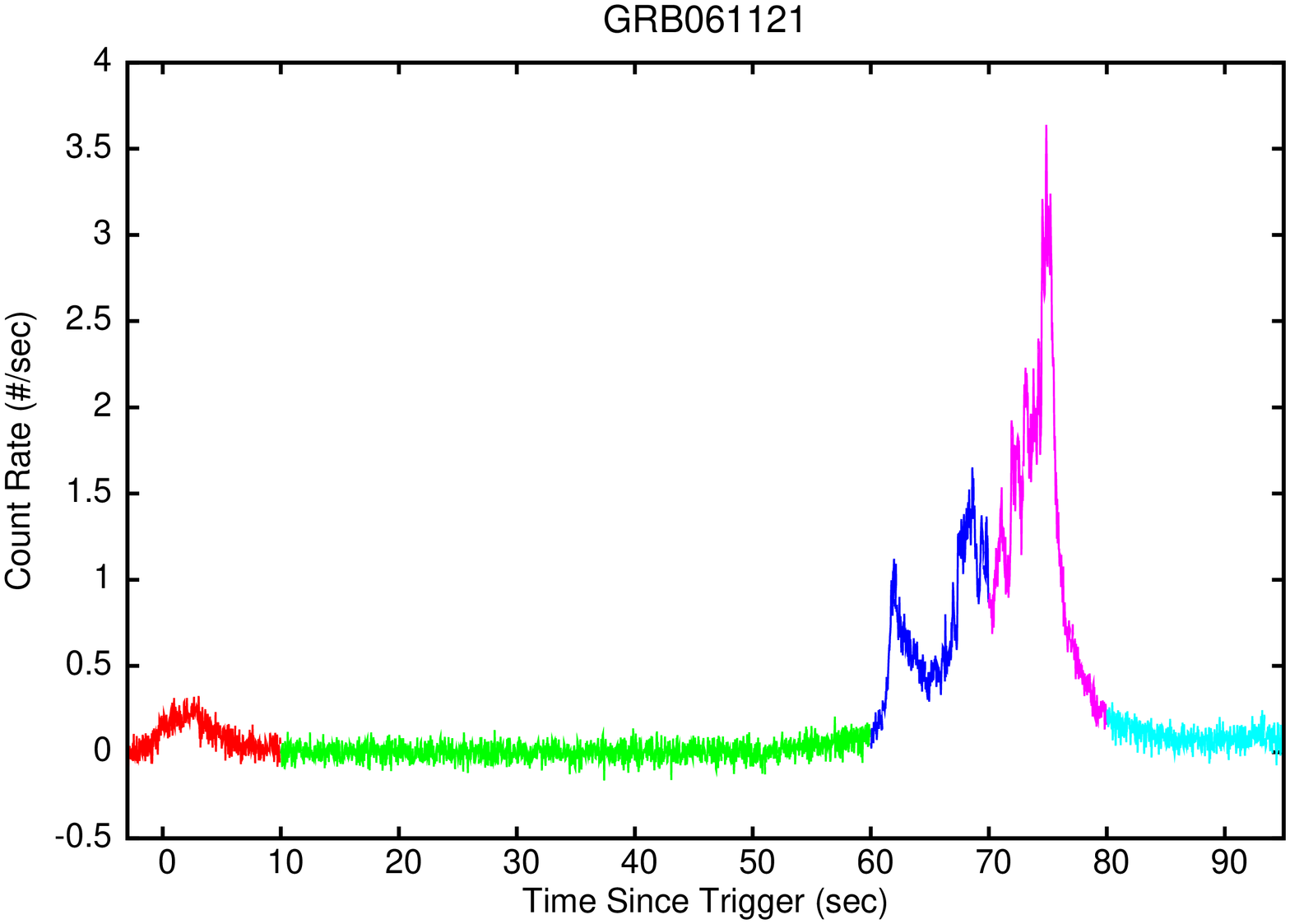}
\includegraphics[width=0.17\textheight, height=0.45\textwidth, angle=-90]{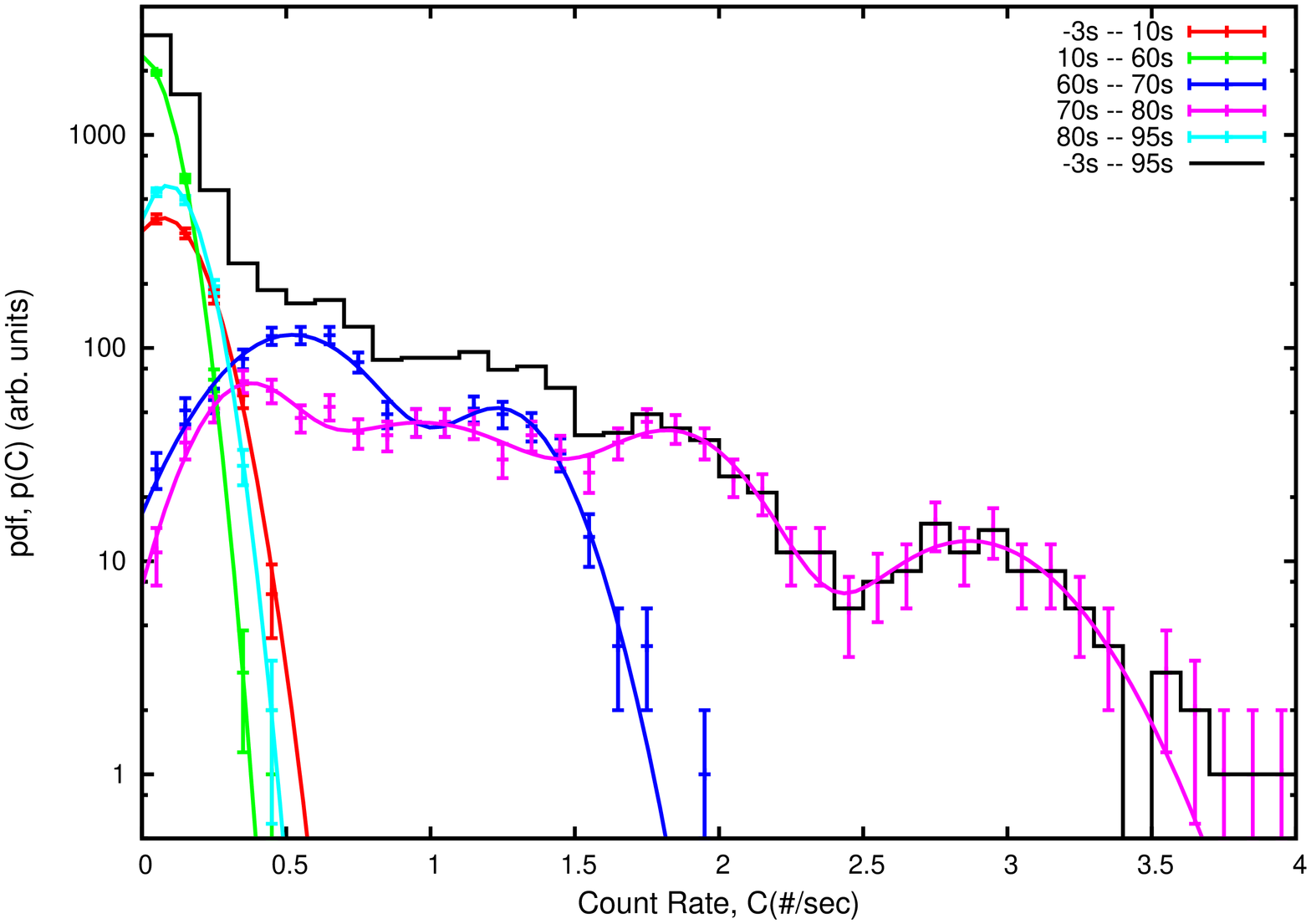}
\includegraphics[width=0.17\textheight, height=0.45\textwidth, angle=-90]{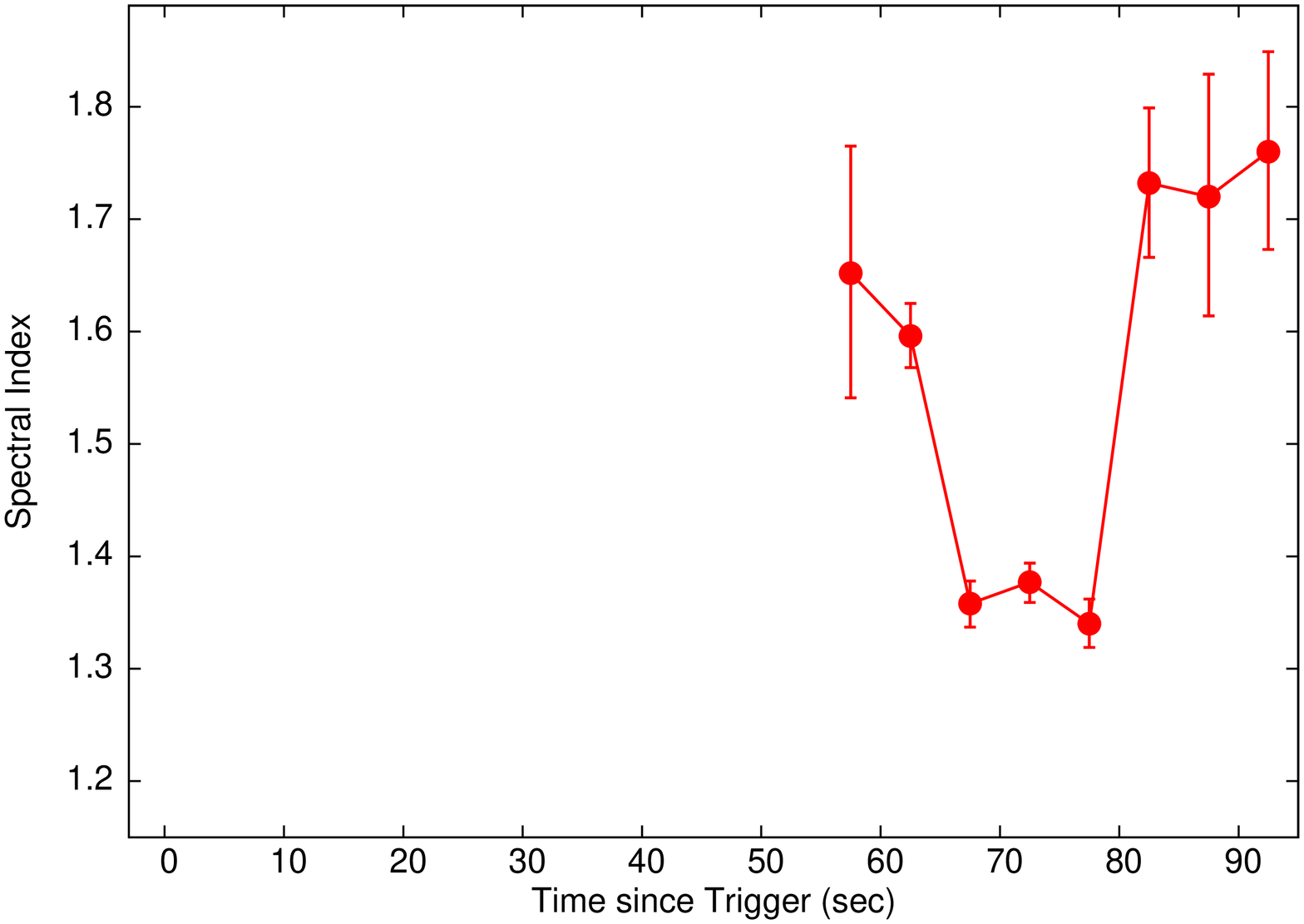}
\includegraphics[width=0.17\textheight, height=0.45\textwidth, angle=-90]{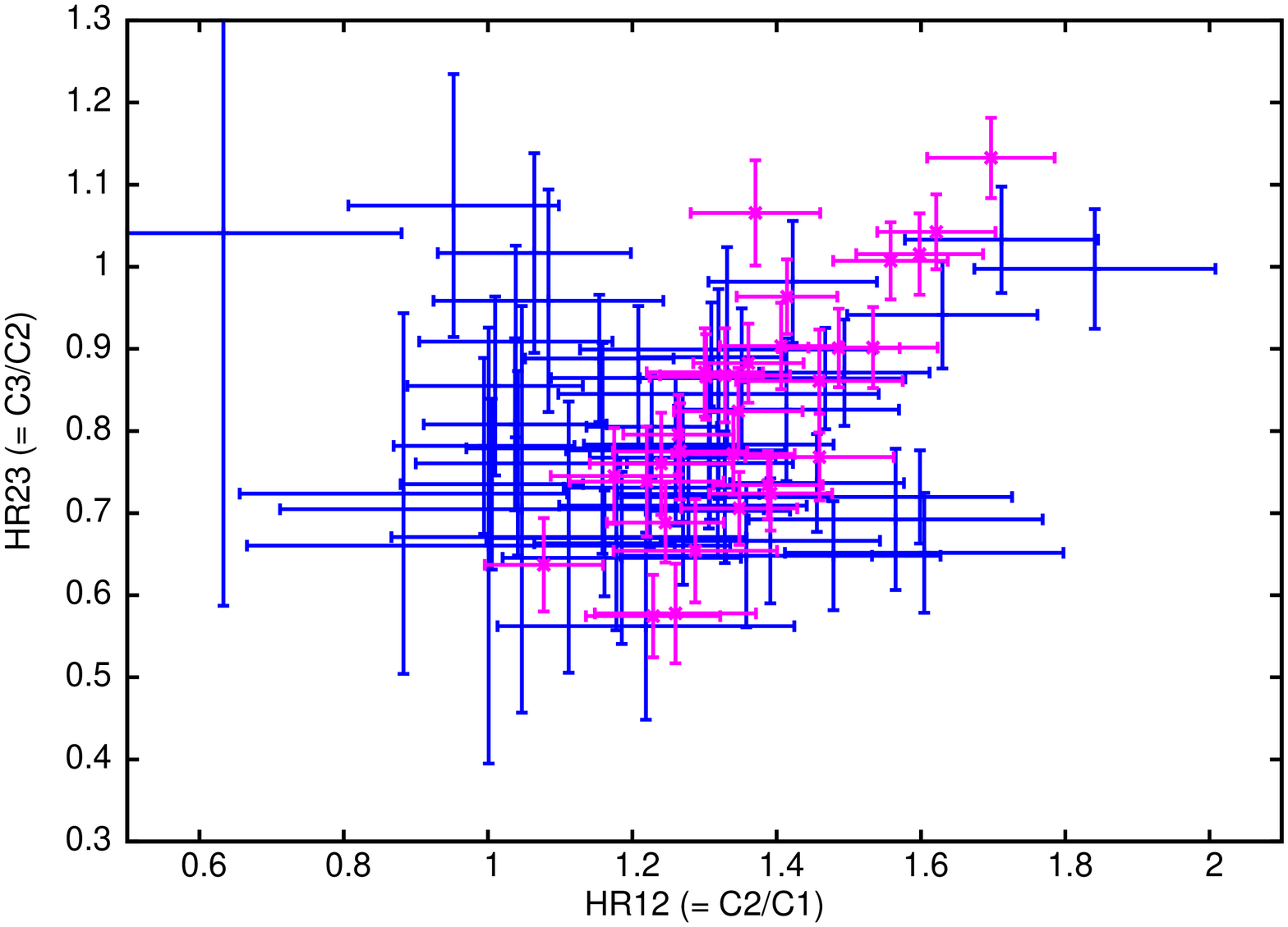}
\caption{Same as Figure \ref{pds1}, but for GRB061121.} \label{pds3}
\end{center}
\end{figure}
\item {\bf GRB080411 :} GRB080411 consists of two strong peaks which are widely separated in time. 
The time averaged spectrum in the {\emph Swift}/BAT energy range was fitted with a power-law 
with spectral index 1.678. The observed fluence is $4.3 \times 10^{-5}$ erg cm$^{-2}$. The measured 
redshift of the burst is 1.03 which gives an isotropic energy of the burst $\sim 4.04 \time 10^{53}$
erg (\cite{GCN080411}).  

\item {\bf GRB080319b :} This gamma-ray burst is one of the brighest GRBs with a very strong optical
flash during its prompt emission. This GRB was simultaneously observed with {\emph Swift}-BAT and 
{\emph Konus}-Wind in the energy range 15--350 keV and 18--1160 keV respectively. The observed fluence in
the 20keV--7MeV band is $5.7 \pm 10^{-4}$  erg cm$^{-2}$. The measured redshift is 0.937 and isotropic 
energy is estimated to be $1.3 \times 10^{54}$ erg. The time averaged spectrum measured by {\emph Konus}-Wind
was fitted by a broken power-law with break energy at $E_{br}=650$ keV. The spectral index below 650 keV 
is $0.18 \pm 0.01$ and above 650 keV is $2.87 \pm 0.44$. This GRB was also followed by other telescopes 
in different wavelengths during the prompt and afterglow emission. During the prompt emission it was
observed by the {\emph TORTORA} telescope in the optical region and it is discussed in detail by 
\cite{Rac08}. 
\begin{figure}
\begin{center}
\includegraphics[width=0.17\textheight, height=0.45\textwidth, angle=-90]{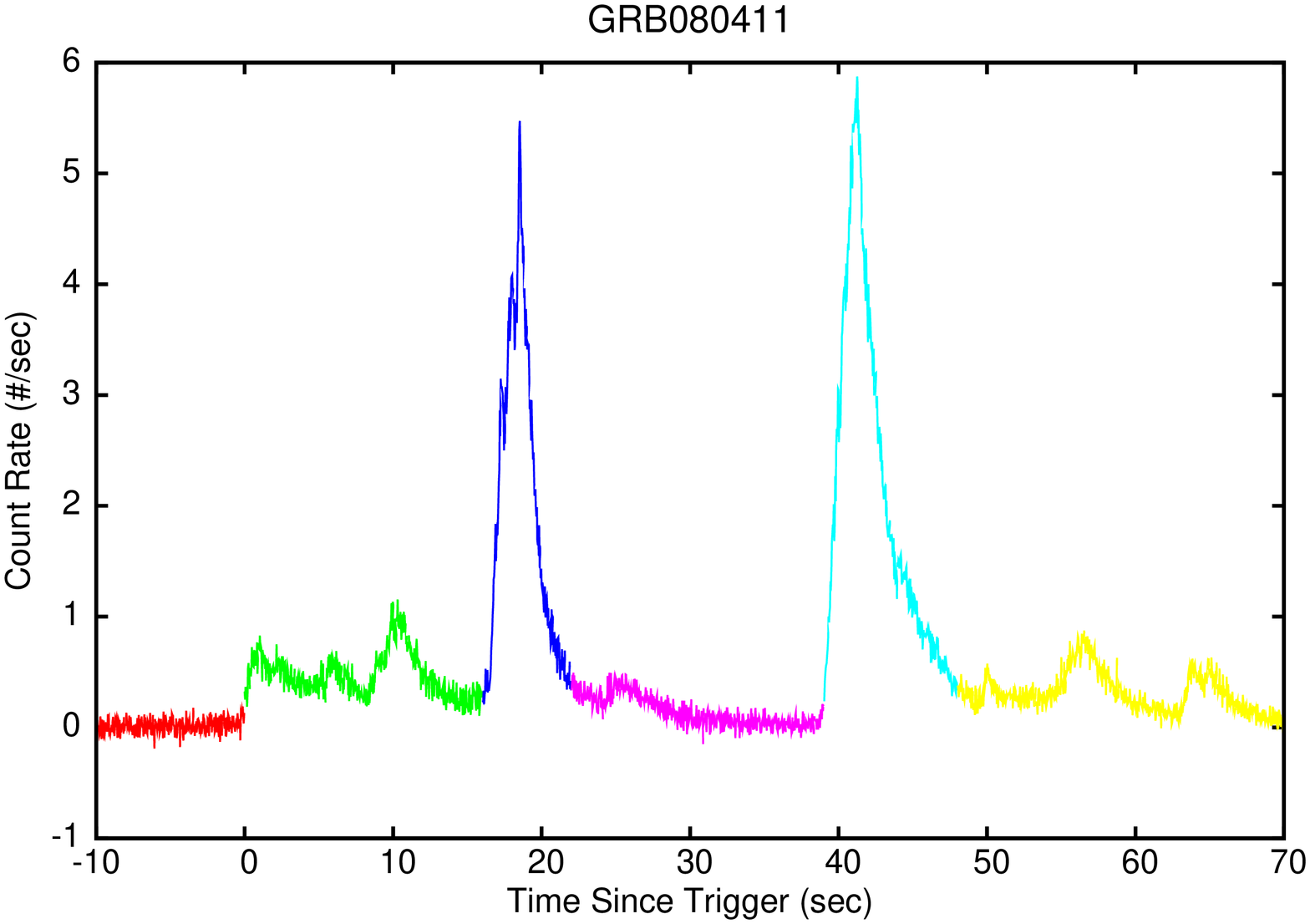}
\includegraphics[width=0.17\textheight, height=0.45\textwidth, angle=-90]{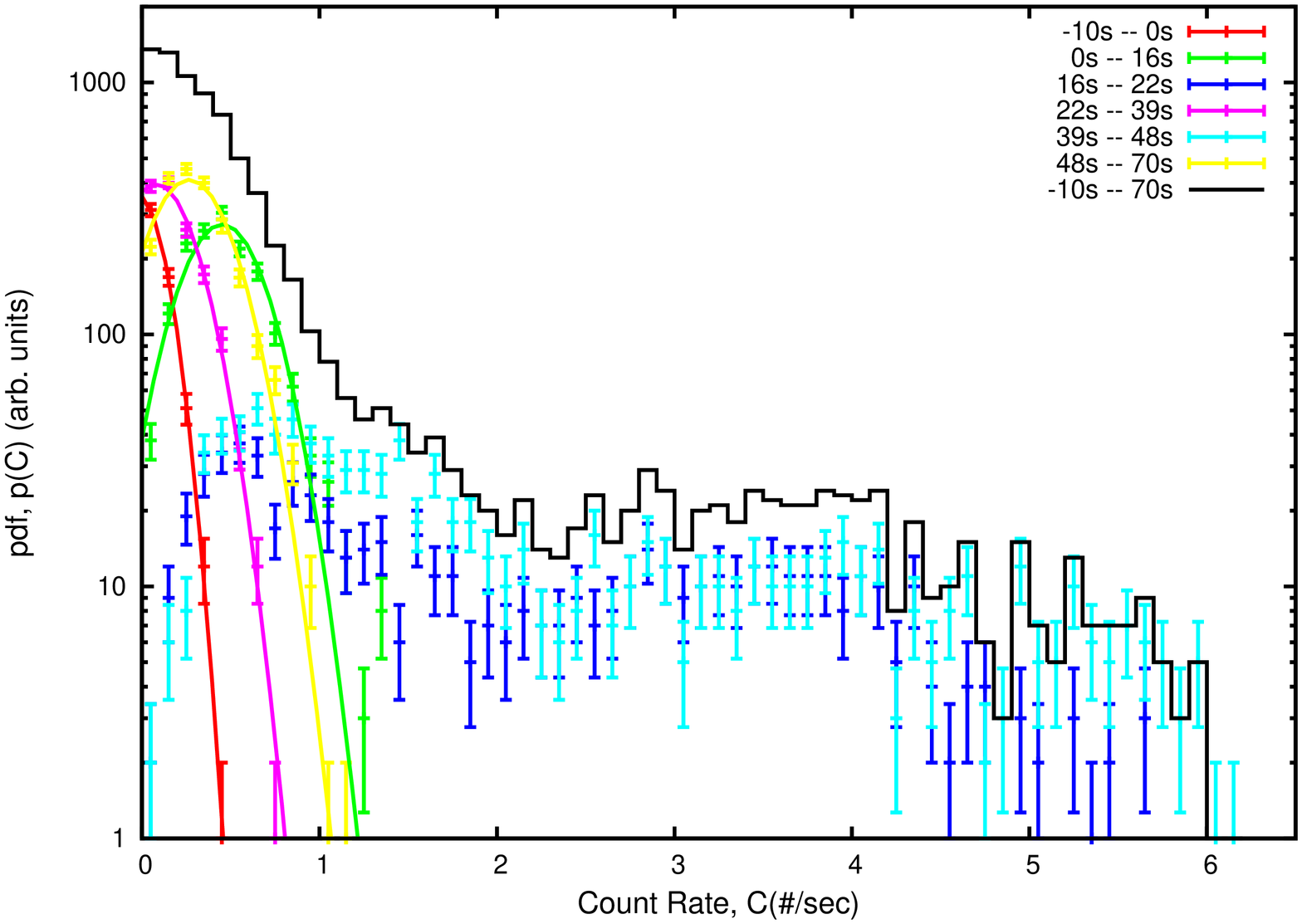}
\includegraphics[width=0.17\textheight, height=0.45\textwidth, angle=-90]{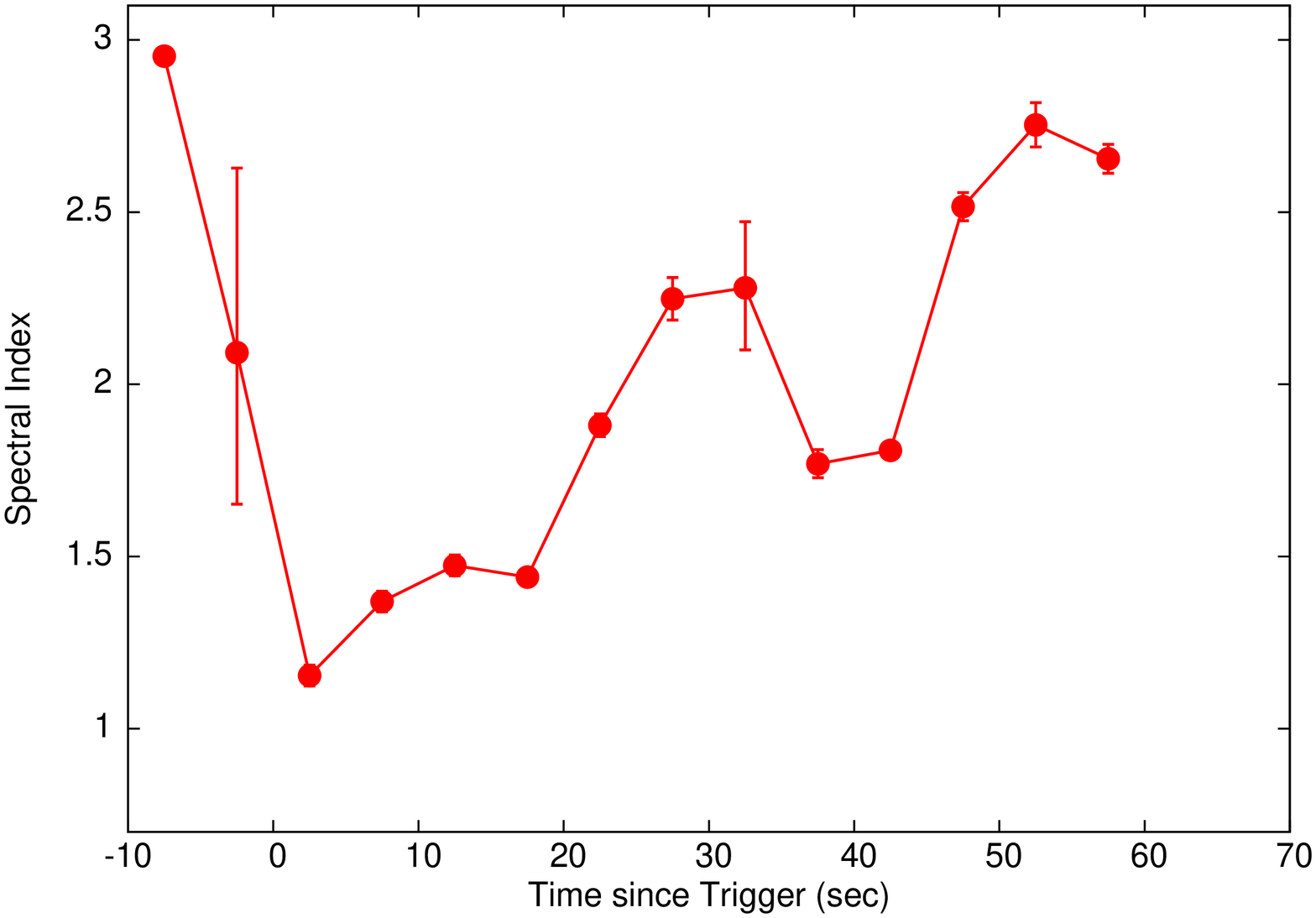}
\includegraphics[width=0.17\textheight, height=0.45\textwidth, angle=-90]{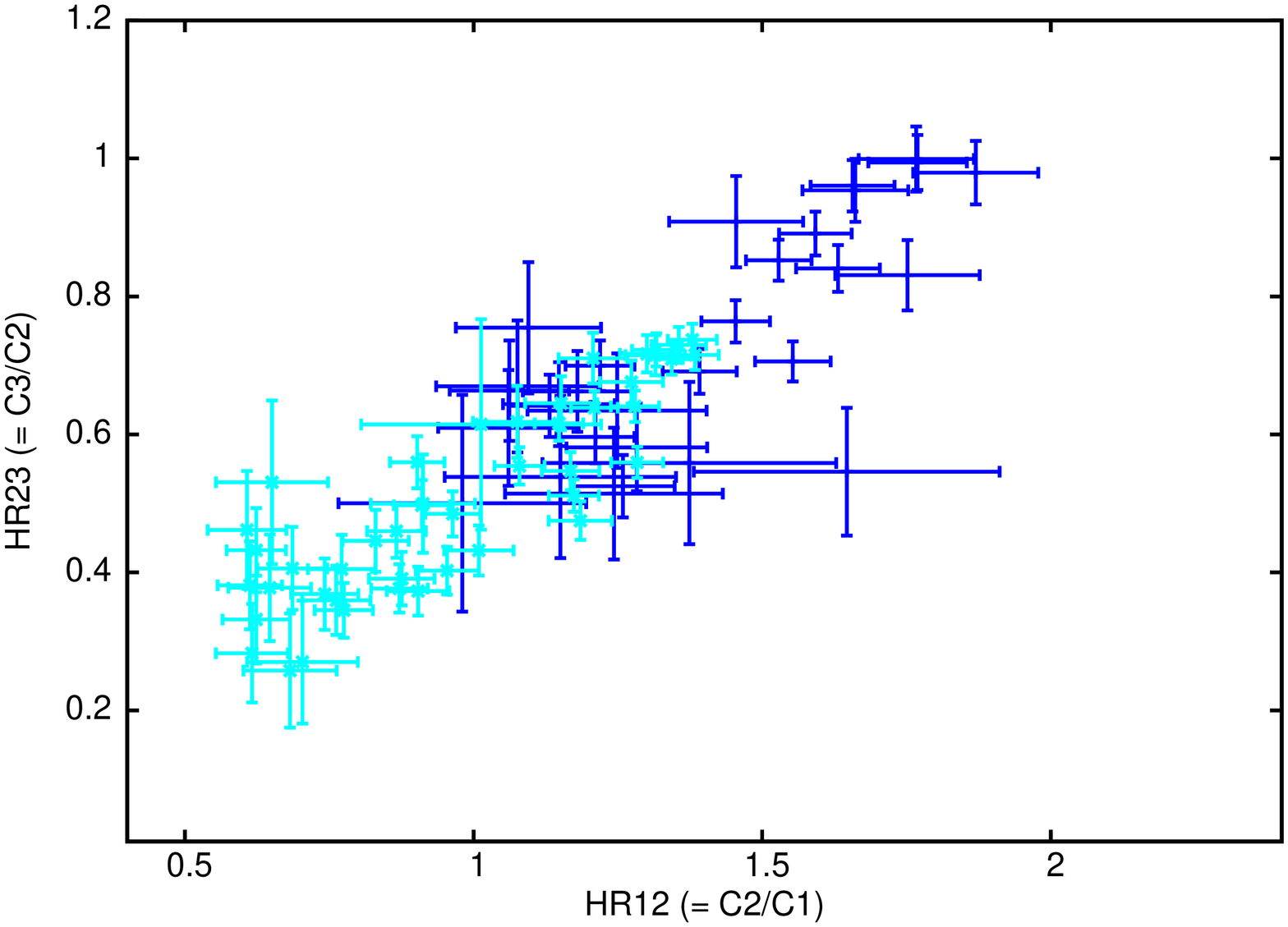}
\caption{Same as Figure \ref{pds1}, but for GRB080411.} \label{pds4}
\end{center}
\end{figure}
\noindent This GRB is studied quite extensively by several authors (\cite{ZFP09, KN09}). The most significant
feature of this particular GRB is that the time averaged spectrum can not be explained by the 
synchrotron-self-Compton (SSC) process in the framework of internal shock model. Rather a physically 
reasonable set of parameters can be obtained in the framework of turbulence model (\cite{KN09}).  
\end{itemize}  

\begin{table}
\begin{center}
\caption{{\bf GRB080319B:} The fitted parameters, their uncertainities and the goodness-of-fit parameter for the Gaussian 
distribution(s) corresponding to each \emph{pdf}. The colours used to plot the \emph{pdf}s are also given in column 1.
For the first time segment, as the \emph{dof} is zero, the fit is a representative one.} \label{tab1}
\begin{tabular}{c|c|c|c|c|c}
\noalign{\hrule height 1pt}
Time & \multicolumn{4}{c|}{Gaussian Parameters} & \multirow{2}{*}{$\chi^2/\nu$} \\ \cline{2-5}
({\it sec}) & $M$ & $A_i$ & $\mu_i$ & $\sigma_i$ & \\
\noalign{\hrule height 1pt}
$-10$ --- $-4$ & 1 & $429.7$ & $-0.07$ & $0.11$ & --  \\ 
(red) & & & & & \\ \hline
$-4$ --- $10$ &\multirow{2}{*}{2}& $93.8\pm4.4$ & $0.66\pm0.03$ & $0.42\pm0.03$ & \multirow{2}{*}{23.16/22}  \\
(green) & & $39.6\pm3.0$ & $1.89\pm0.06$ & $0.40\pm0.04$ &  \\ \hline
$10$ --- $30$ & 1 & $170.1\pm4.6$ & $2.64\pm0.01$ & $0.44\pm0.01$ & 17.13/20  \\ 
(blue) & & & & & \\ \hline
$30$ --- $52$ &\multirow{3}{*}{3}& $26.5\pm8.6$ & $1.02\pm0.17$ & $0.23\pm0.09$ & \multirow{3}{*}{27.46/18}  \\
(magenta) & & $59.5\pm25.8$ & $1.57\pm0.08$ & $0.21\pm0.1$ & \\
& & $184.8\pm11.6$ & $2.16\pm0.05$ & $0.36\pm0.03$ &   \\ \hline
$52$ --- $60$ &\multirow{2}{*}{2} & $235.9\pm44.9$ & $0.23\pm0.004$ & $0.11\pm0.01$ & \multirow{2}{*}{0.85/3}  \\
(cyan) & & $24.1\pm24.0$ & $0.43\pm0.34$ & $0.21\pm0.13$ &  \\
\noalign{\hrule height 1pt}
\end{tabular}
\end{center}
\end{table}

\begin{table}
\begin{center}
\caption{{\bf GRB050525:} Same as Table \ref{tab1}
For the last time segment, as the \emph{dof} is zero, the fit is a representative one.} \label{tab2}
\begin{tabular}{c|c|c|c|c|c}
\noalign{\hrule height 1pt}
Time  & \multicolumn{4}{c|}{Gaussian Parameters} & \multirow{2}{*}{$\chi^2/\nu$} \\ \cline{2-5}
({\it sec}) & $M$ & $A_i$ & $\mu_i$ & $\sigma_i$ & \\
\noalign{\hrule height 1pt}
$-3$ --- $0$  & 1 & $96.5\pm9.5$ & $0.043\pm0.027$ & $0.086\pm0.014$ & $0.7/1$  \\
(red) & & & & & \\ \hline
$0$  --- $4$  &  & $20.6\pm2.2$ & $1.23\pm0.03$   & $0.32\pm0.03$   &   \\
(green) & 3 & $6.3\pm1.4$  & $2.37\pm0.08$   & $0.28\pm0.08$   &  40.8/37 \\
& & $5.9\pm1.0$  & $3.82\pm0.09$   & $0.54\pm0.09$   &   \\ \hline
$4$  --- $12$ & & $34.0\pm3.5$ & $0.26\pm0.02$   & $0.19\pm0.03$   & \multirow{3}{*}{28.5/41}  \\
(blue) & 3 & $29.9\pm1.8$ & $1.24\pm0.04$   & $0.48\pm0.04$   &   \\
& & $17.9\pm1.2$ & $3.59\pm0.03$   & $0.57\pm0.03$   &   \\ \hline
$12$ --- $15$ & 1 & $113.9$      & $0.001$         & $0.02$ &  -- \\
(magenta) & & & & & \\ 
\noalign{\hrule height 1pt}
\end{tabular}
\end{center}
\end{table}

\section{Data Analysis}
We analysed {\emph Swift}/BAT event data for four {\emph Swift} detected gamma -ray bursts GRB050525, 
GRB061121, GRB080411 and GRB080319b. The standard BAT software and the calibration database 
(CALDB : 20090130) have been used to obtain the lightcurves and the spectra for all four GRBs. 

\noindent We extracted lightcurves with 10 ms time resolution in the energy band 15--150 keV for all GRBs. 
We also obtained lightcurves in the energy bands $E_1$ : 15--25 keV,
$E_2$ : 25--50 keV and $E3$ : 50--100 keV with time resolution of 200 msec and 1 sec for all GRBs.   

\noindent The spectra are extracted for every 5 sec from -10 sec
with respect to the trigger time for GRB061121, GRB080411 and GRB080319b. In case of GRB050525 spectra 
are extracted for every 2.5 sec. 

\section{Results and Discussion}
\subsection{Probability density function}
A stochastic process $\{x(t)\}$, also known as time series, is an ensemble of real or complex valued functions.
Each stochastic process is characterized by a probability density function (\emph{pdf}) depending on the 
underlying physical process 
responsible for the generation of the time series (\cite{BP10}). Therefore it is always interesting to develop an understanding 
of the probability density function for a given time series. 

\noindent A time series can be stationary or non-stationary depending on whether the statistical properties 
of the time series, such as mean, variance, are independent of time or not. The gamma-ray burst time series
is a real non-stationary time series with statistical properties of the time series vary with time. Here we 
calculate the time evolution of the \emph{pdf} and the time averaged \emph{pdf} for all four GRBs of our sample.
To study the time evolution of the \emph{pdf} we divide the GRB time series into different time segments and
for each segment we calculate the \emph{pdf}. If $\tau$ be the time length of one segment starting at time t 
then the \emph{pdf} is defined by 
\begin{equation}
p_{t,t+\tau} (x) = \frac{\Delta N_x}{N} 
\end{equation}
where $\Delta N_x$ the number of trials that the random variable falls between $x$ 
and $x+\Delta x$, $N$ is the total number of trials. The error in $p(x)$ at $x$ is
given by $\sqrt{\Delta N_x}$. For GRB time series $x$ represents the {\emph count rate}.  
The length of the time segment $\tau$ at time t is chosen through visual inspection
of the GRB time structure. For the time-averaged \emph{pdf}, $p(x)$ is computed for
the complete lightcurve. It is to be noted that in this work we have not used the normalization for
$p(x)$. The time-averaged \emph{pdf}, \emph{pdf}s for different time
segments for each GRB, alongwith their respective lightcurves,  are shown in Figures \ref{pds1}--\ref{pds4} 
(\emph{first} and \emph{second} panel).
The different portions of the lightcurve are shown in different colours and their corresponding 
probability density functions are also given with the same colour. The time resolution
of the lightcurve used for the calculation is 10 ms for all GRBs. From the \emph{pdf}s plotted for
different time segment for each GRB it is evident that the \emph{pdf}s corresponding to
the rising and the falling part of GRBs peak around zero and those are not affected by the other
portions of the GRB. As the GRB evolves, the \emph{pdf} also evolves. In almost all cases, the \emph{pdf}
for each time segment is fitted with single or multiple Gaussian distributions given by
\begin{equation}
F(x) = \sum\limits_{i=1}^{M} A_{i}\,\exp\left[-\frac{1}{2}\left(\frac{x-\mu_i}{\sigma_i}\right)^{\!\!2}\right]
\end{equation}
where $M\geq1$. For all GRBs, the parameters of the fitted Gaussian 
function(s), their uncertainities and the goodness-of-fit parameter are listed in Tables \ref{tab1}--\ref{tab4}. 
For the time segments before the GRB trigger and after the GRB episode, the lightcurves are dominated
by Poisson noise. Therefore the number of points obtained in the \emph{pdf} for these time segments are
small. This results in very poor fit statistics. The variation of the parameters of the Gaussian
distributions with the evolution of the burst reflects the non-stationarity of the underlying random process. We also
attempted to fit the \emph{pdf}s with log-normal functions but the fitting with Gaussian function was found to be 
statistically better than the log-normal distribution.
There is no definite correlation between the peak position and width 
of the Gaussians. It is very clear that the time-averaged probability density function profile (represented 
in \emph{black}) can be fitted with multiple Gaussian functions. The 
possible physical interpretation of the Gaussian distribution is discussed below. It is argued that
the Gaussian distribution possibly indicates the presence of turbulence in the GRB outflow originated
in a relativistically expanding fireball. Possible consequences of the turbulence in the spectral 
evolution of each of the selected GRBs are also discussed in detail.   
\begin{table}
\begin{center}
\caption{{\bf GRB061121:} Same as Table \ref{tab1}} \label{tab3}
\begin{tabular}{c|c|c|c|c|c}
\noalign{\hrule height 1pt}
Time & \multicolumn{4}{c|}{Gaussian Parameters} & \multirow{2}{*}{$\chi^2/\nu$} \\ \cline{2-5}
({\it sec}) & $M$ & $A_i$ & $\mu_i$ & $\sigma_i$ & \\
\noalign{\hrule height 1pt}
$-3$ --- $10$ & 1 & $408.1\pm14.4$ & $0.07\pm0.01$ & $0.14\pm0.01$ & 1.6/2  \\ 
(red) & & & & & \\ \hline
$10$ --- $60$ & 1 & $2374\pm144$ & $-0.01\pm0.01$ & $0.1\pm0.004$ & 0.9/2  \\ 
(green) & & & & & \\ \hline
$60$ --- $70$ & 2 & $115.4\pm5.5$ & $0.52\pm0.01$ & $0.26\pm0.01$ & 10.3/12  \\ 
(blue) & & $49.5\pm4.2$  & $1.25\pm0.02$  &  $0.18\pm0.02$ & \\ \hline
& \multirow{4}{*}{4} & $54.8\pm13.5$ & $0.35\pm0.02$ & $0.16\pm0.03$ & \multirow{4}{*}{12.5/22}  \\
$70$ --- $80$ & & $44.6\pm2.8$ & $0.97\pm0.07$ & $0.39\pm0.14$ &  \\
(magenta) & & $37.2\pm6.1$ & $1.85\pm0.05$ & $0.24\pm0.04$ & \\
& & $12.4\pm1.4$ & $2.87\pm0.04$ & $0.32\pm0.05$ & \\ \hline
80 --- 95 & 1 & $581.0\pm6.9$ & $0.09\pm0.002$ & $0.10\pm0.002$ & 0.3/2  \\
(cyan) & & & & &\\ 
\noalign{\hrule height 1pt}
\end{tabular}
\end{center}
\end{table}

\subsection{\emph{pdf} and Turbulence}
It is shown by \cite{KN09} and \cite{ZFP09} that the optical and gamma-ray spectra
in the prompt emission of GRB080319B can not be explained in the framework of internal shock model. 
Instead, \cite{NK09} proposed the relativistic turbulence model to explain the prompt
emission spectra of GRB080319B. In the framework of this model the highly variable gamma-ray lightcurve
can be produced due to the emission from randomly moving eddies in the turbulence. The random 
relativistic variations of velocities of eddies cause the large amplitude variations in the individual 
pulses in gamma-ray burst lightcurve. In case of GRB080319B the optical emission is produced by 
synchrotron process in the inter-eddy medium whereas the gamma-rays are produced by inverse Compton 
process by energetic electrons in the eddies in the turbulence. To understand the probability density 
function obtained for the X-ray lightcurve of GRB080319B we argue as follows. 
\begin{table}
\begin{center}
\caption{{\bf GRB080411:} Same as Table \ref{tab1}} \label{tab4}
\begin{tabular}{c|c|c|c|c|c}
\noalign{\hrule height 1pt}
Time & \multicolumn{4}{c|}{Gaussian Parameters} & \multirow{2}{*}{$\chi^2/\nu$} \\ \cline{2-5}
({\it sec}) & $M$ & $A_i$ & $\mu_i$ & $\sigma_i$ & \\
\noalign{\hrule height 1pt}
$-10$ --- $0$ & 1 & $359.2\pm26.8$ & $-0.02\pm0.02$ & $0.14\pm0.01$ & 0.4/2  \\ 
(red) & & & & & \\ \hline
$10$ --- $16$ & 1 & $293.9\pm18.7$ & $0.43\pm0.02$ & $0.20\pm0.01$ & 17.0/5  \\ 
(green) & & & & & \\ \hline
$22$ --- $39$ & 1 & $396.4\pm19.1$ & $0.08\pm0.02$ & $0.21\pm0.01$ & 8.3/5  \\ 
(magenta) & & & & &\\ \hline
$48$ --- $70$ & 1 & $442.3\pm23.5$ & $0.27\pm0.01$ & $0.20\pm0.01$ & 12.8/4  \\ 
(yellow) & & & & &\\ 
\noalign{\hrule height 1pt}
\end{tabular}
\end{center}
\end{table}

\noindent If the large amplitude variations are due to the random velocity fluctuations in turbulent 
eddies then the statistical properties of the velocity fluctuations are expected to be translated into
the random fluctuations of the lightcurves. In recent works by \cite{M+02} and \cite{C+96},
it has been observationally shown that the probability density function of turbulent velocity fluctuations
is Gaussian for fully developed turbulence. The probability density function is sub-Gaussian  when
turbulence is developing and it is hyper-Gaussian during the decay of turbulence. We assume that 
these properties are also satisfied by the random velocity fluctuations in relativistic turbulence.
In such a condition the intensity fluctuations in GRB lightcurve (which are supposed to be due to the 
velocity fluctuations of turbulent eddies) should also follow a Gaussian probability density function.
In Figures \ref{pds1}--\ref{pds4} (\emph{second panel}) the \emph{pdf}s are shown for GRB080319b, GRB050525,
GRB061121 and GRB080411 respectively. The \emph{pdf}s for different time segments are shown in 
different colour and the time-averaged \emph{pdf} is represented in \emph{black}. The colour sequence is maintained same in 
all GRBs. It is to be noticed that the \emph{pdf}s corresponding to the rising and falling edges of GRBs are completely
separated from other \emph{pdf}s corresponding to different time segments. In case of GRB080319b the \emph{pdf}
corresponding to a single segment is fitted with a single Gaussian function whereas for GRB050525 and
GRB061121 \emph{pdf}s corresponding to a single segment are fitted with multiple Gaussian functions. Multiple Gaussian 
distributions corresponding to a single lightcurve segment possibly indicate more than one region 
of turbulence in the GRB outflow. If we compare the rising and falling portions of GRBs with the development and decay of
turbulence respectively then the corresponding peaks could be sub- and hyper-Gaussian distributions respectively. But
such conclusion is very difficult to make with the present data.   

\noindent At this point it is important to mention that in presence of intermittency, the turbulence 
statistics deviates from Gaussian behaviour (\cite{FRL10, V95}). Generally when the amplitude of fluctuations are small, 
the turbulence statistics is Gaussian in case of fluid turbulence. In case of magnetohydrodynamic 
turbulence intermittency is generally present. The origin and nature of intermittency in fluid as well
as magnetohydrodynamic turbulence is not known. Therefore any departure of the
peaks in observed probability density functions from Gaussian distribution does not rule out the present
interpretation of \emph{pdf}. Prticularly, if we consider the \emph{pdf} of GRB080411 (FIgure \ref{pds4}),
we can see that the \emph{pdf}s corresponding to the \emph{red, green, magenta} and \emph{yellow} portions of the 
lightcurve can be fitted with one or more Gaussian functions, but the \emph{pdf}s corresponding to the
peaks (\emph{blue} and \emph{cyan}) can not be fitted with Gaussian functions at all. It appears more like a power-law.
Here we have  not deliberately fitted them with any functional form. 

\noindent Having argued that the \emph{pdf}s actually reflect the turbulence origin of GRB, now we concentrate on the spectral
properties of GRB to see if those properties can be explained in terms of turbulence. 

\subsection{Turbulence, particle acceleraion and spectral evolution of GRBs}
\subsubsection{Turbulence and particle acceleration}
The major challenge in high energy astrophysics is to understand the process of accelerating charged
particles to relativistic energies so that they can emit high energy radiation in a given physical condition.   
It is generally assumed that the first order Fermi acceleration across a shock (internal or external) 
accelerates particles to relativistic energies with a non-thermal distribution. But GRB models in
general do not discuss this issue in detail. In presence of turbulence which is presumably magnetohydrodynamic
in GRB outflow the particles can be accelerated by stochastic acceleration process or by magnetic reconnection
process. In case of stochastic acceleration particles interact with the MHD waves developed in the 
plasma flow. This is a second order Fermi process and it is generally less efficient compared to the shock
acceleration process. \cite{Laz03} discussed that the stochastic acceleration process can be more
efficient if momentum diffusion rate is higher than the pitch angle scattering rate and that is possible for
a low density plasma in presence of high magnetic field. In such a condition the Alfven speed (in the units
of \emph{c}) exceeds unity and such conditions exist in GRB outflow (\cite{Laz03}). \cite{Mao11} 
discussed jitter radiation in small scale random magnetic field generated by turbulence. Following \cite{HH05},
 they considered that particles are accelerated in small scale random magnetic field. In such a condition,
particle and photon spectra depend on the energy spectrum of turbulence. 

\noindent Otherwise also 
particle acceleration in presence of turbulence is possible through magnetic reconnection which is obvious in
MHD turbulence. Turbulence possibly accelerates the reconnection process. \cite{LV99} first 
put forward the turbulence reconnecton model and \cite{GDPL03} discussed the particle
acceleration in turbulent reconnection. In general particles can be accelerated almost instantly by magnetic
reconnection process, but this does not generate a power-law spectrum of particles. But \cite{GDPL03} 
and \cite{Laz10} showed that the particle acceleration in turbulent reconnection is a first order Fermi process 
and a particle's energy changes after each crossing of the reconnection zone. The average fractional gain in 
energy is proportional to the reconnection speed. The magnetic flux from two neighbouring regions of the fluid
approach each other with the reconnection speed. It was shown by \cite{GDPL03} that the
resultant particle spectrum is a power-law with spectral index -2.5. This is steeper than the spectrum ($\gamma^{-2}$)
obtained with shock acceleration process, $\gamma$ being the particle Lorentz factor. 
\cite{Lar03} also studied the lepton acceleration by relativistic
collisionless magnetic reconnection. They obtained a distribution function for electrons which varies as $\sim \gamma^{-1}$,
but the final spectrum (what the authors called apparent spectum) depends on the distribution function of the
maximum Lorentz factor of electrons in each reconnection event. This makes the electron distribution $\sim \gamma^{-1-q}$
where $q > 1$.  
\subsubsection{Spectral properties}
We study the following spectral properties of GRBs of our sample to understand if turbulence can explain the observed
results.
\begin{itemize} 
\item {\bf Spectral evolution :} To study the spectral evolution of selected GRBs we extract the spectrum for 
every 5 sec for GRB080319B,
GRB061121 and GRB080411 and for every 2.5 sec for GRB050525. For the first three GRBs, the photon spectra 
in the 15--150 keV energy range are fitted with power-law ($\sim\epsilon^{-\alpha}$), 
whereas for the last one the spectra are fitted with a power-law with exponetial cut-off. The spectra are
fitted using $\chi^2$ minimization process within XSPEC. 

\noindent For GRB080319b, GRB061121 and GRB080411, the magnitude of spectral index grossly increases as the
burst evolves, indicating that the spectrum evolves from hard--to--soft with time. This is indeed observed for
other bursts reported in the literature. The value of photon spectral index varies from $\sim$0.6 to $\sim$2.1
for GRB080319b, from $\sim$1.3 to $\sim$1.8 for GRB061121 and from $\sim$1.1 to $\sim$2.6 for GRB080411. 
It is to be noted that the evolution of spectral index is model dependent. Here the spectra are fitted with a 
single power-law which is statistically adequate due to the narrow energy bandwidth of the detector. It was shown by
\cite{Band93} that the GRB spectra, over a broader energy range are well-fittted with a smoothly broken spectrum (Band spectrum). 
The observed spectral index variation reflects the evolution of underlying GRB spectrum which is intrinsically
a smoothly broken spectrum with different spectral slopes on two sides of the break energy. If we
consider the turbulent reconnection responsible for the acceleration of particles to relativistic energies and
synchrotron process is the basic radiation mechanism then the value of photon spectral index should be around $1.75$ 
for instantaneous spectrum and $2.25$ for cooling dominated spectrum. But the value of observed photon spectrum indices
fall below these limits. This implies that even if turbulent reconnection plays an important role in accelerating particles
in GRB outflow, it is not the only mechanism working there. This requires further detailed study.

\noindent In case of GRB050525, the spectral evolution pattern is not very clear because of its short duration. The spectra of 
this GRB could not be fitted well with power-law, instead a power-law with exponential cut-off fitted the spectra well.

\begin{figure}
\begin{center}
\includegraphics[width=0.12\textheight,angle=-90]{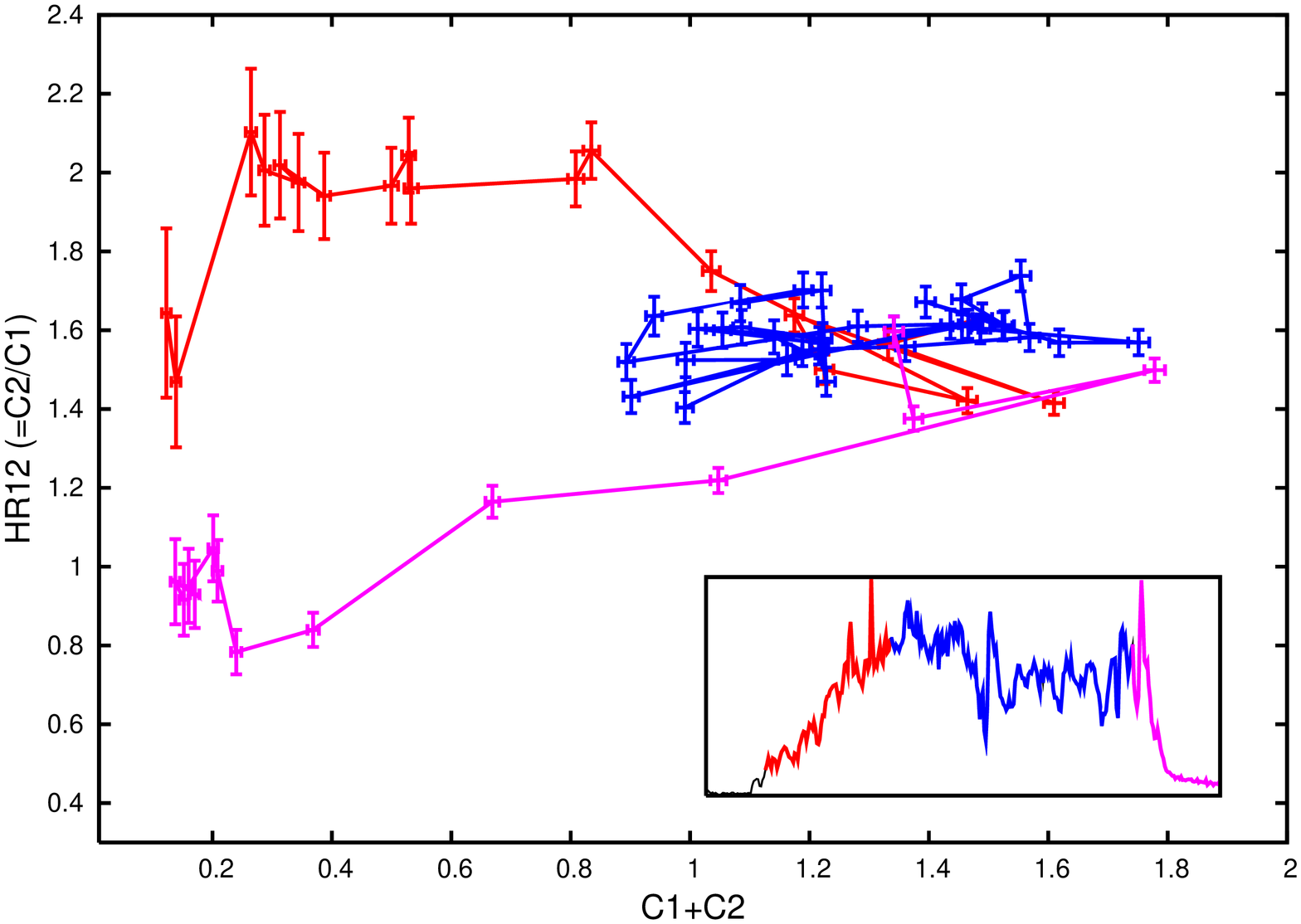}
\includegraphics[width=0.12\textheight,angle=-90]{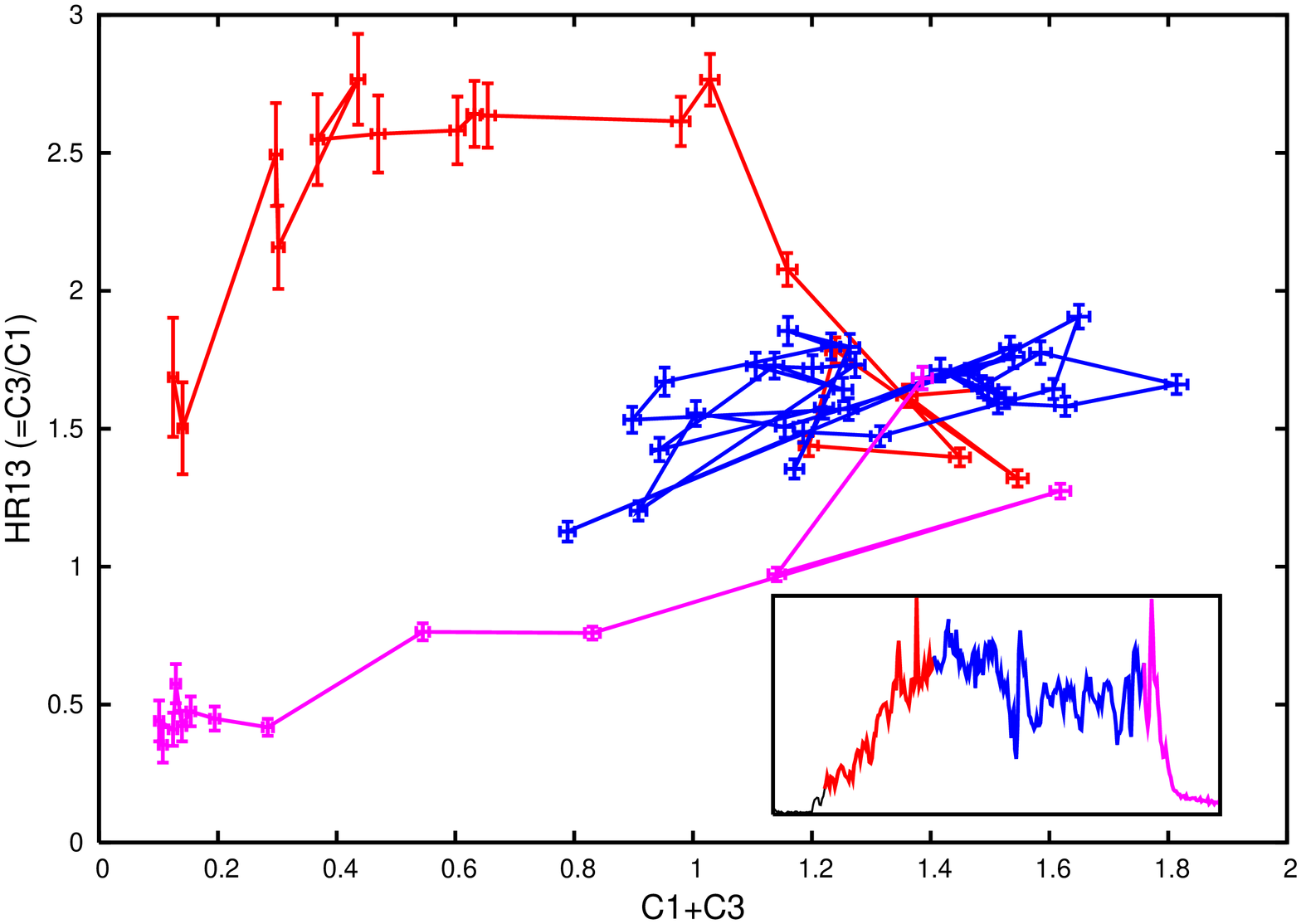}
\caption{GRB080319B: \emph{Left panel :} Hardness -- count-rate diagram for energy bands $E_1$(15--25 keV) and $E_2$(25--50 keV),
\emph{Right panel :} Hardness -- count-rate diagram for energy bands $E_1$(15--25 keV) and $E_3$(50--100 keV). The colours used
for different portions of the lightcurves (plotted as inset in each panel) are in direct correspondence with the colours used
in hardness -- count-rate diagram.} \label{hr1}
\end{center}
\end{figure}

\begin{figure}
\begin{center}
\includegraphics[width=0.12\textheight,angle=-90]{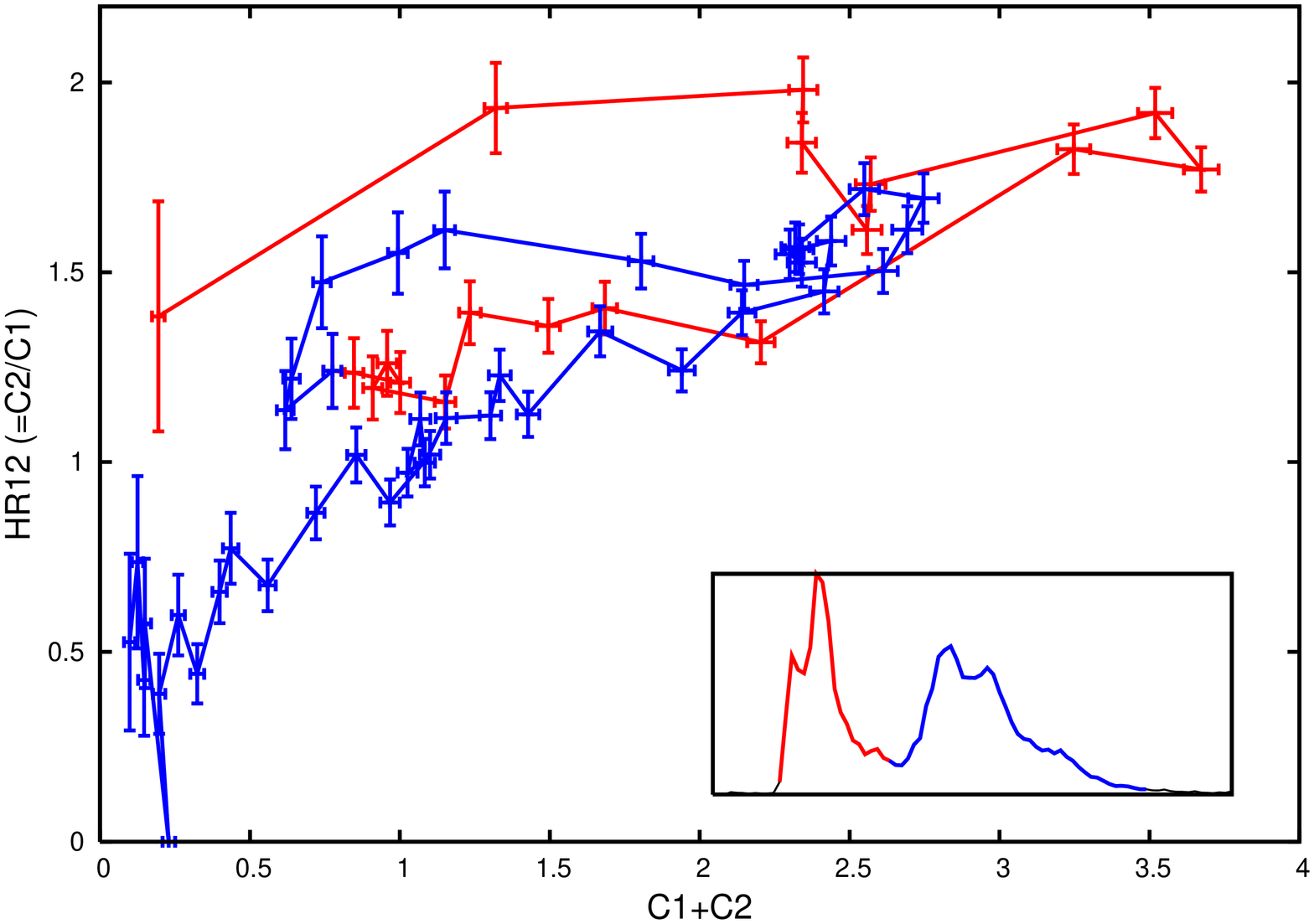}
\includegraphics[width=0.12\textheight,angle=-90]{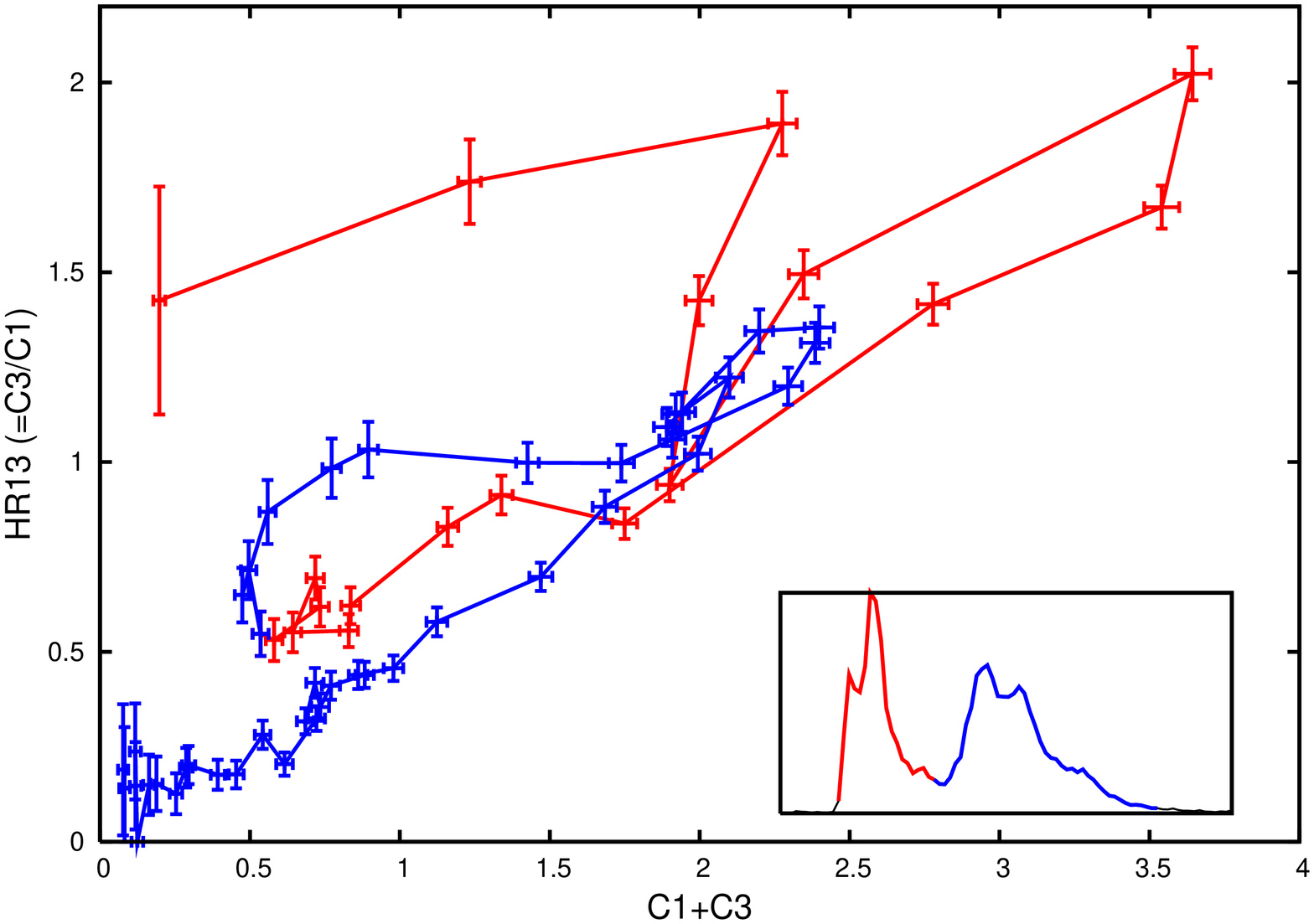}
\caption{GRB050525: same as Figure \ref{hr1}.} \label{hr2}
\end{center}
\end{figure}

\item {\bf Colour-colour diagram :} The hardness ratio between the two energy bands, say $E_1$ and $E_2$, is defined as 
\begin{equation}
HR_{12} = \frac{C_1}{C_2}
\end{equation}
where $C_1$ and $C_2$ are the count rates in the energy bands $E_1$ and $E_2$ respectively. The error in the 
estimation of hardness ratio is
\begin{equation}
\Delta HR_{12} = \sqrt {\left(\frac{\Delta C_1}{C_1}\right)^2 + \left(\frac{\Delta C_2}{C_2}\right)^2}
\end{equation}
Similarly the hardness ratio $HR_{13}$ and its error $\Delta HR_{13}$ can be defined for the energy
bands $E_1$ and $E_3$. 

\noindent Colour-colour diagrams, shown in Figure \ref{pds1}--\ref{pds4} (\emph{fourth panel}), show the correlations between the hardness
$HR_{12}$ and $HR_{23}$. These 
plots also have similar colour sequence as the lightcurves and are generated for the active part of the GRB lightcurvers.
The time resolutions used for these plots are 1 sec (GRB080319b, GRB080411) and 200 msec (GRB050525, GRB061121). 
It is generally found that the colour-colour diagrams show  systematic correlations  over the complete evolution of all bursts.
The striking feature for the colour-colour diagrams for all GRBs is that they start with a very hard spectrum. This is possibly
due to the almost instantaneous acceleration of particles to high energies. This could be possible in magnetic current filaments
generated in turbulence. The estimated acceleration time scale for such event is given by \cite{Mao11}. It depends on magnetic
field strength and turbulent length scale in the flow. Later part of the evolution in colour-colour plane is primarily dominated
by the cooling of particles. This makes the particle spectrum steeper and so the photon spectrum. It is important to note that in
case of GRB080319b, the blue and magenta points (Figure \ref{pds1}, \emph{fourth panel}) in colour-colour plane are clustered in a small 
region. These points correspond to the nearly flat portion of the lightcurve (represented in same colours). During this portion
particle acceleration rate and cooling rate were comparable, thereby making the hardness almost constant.

\item {\bf Hardness--Count rate correlation :}  The hardness and count rate correlations have been studied for the active phase of each GRB for the
energy bands $E_1$(15--25keV), $E_2$(25--50keV) and $E_3$(50--100keV). For each burst, hardness vs countrate plots are
shown in Figures \ref{hr1} -- \ref{hr4}. The left panel in each figure shows the hardness vs count rates calculated for energy
bands $E_1$ and $E_2$ while the right panel shows the same for energy bands $E_1$ and $E_3$. The representative colours in the lightcurve show the 
corresponding portions mapped into the hardness-count rate plane. The lightcurves used for this 
calculation had a time resolution of 200 ms except for GRB080319B where the lightcurves had time resolution
of 1 sec. All hardness-count rate loops for GRB080319B, GRB080411 and GRB050525 are in clockwise sense except
for the burst GRB061121. For GRB061121 the loop structure is quite complex. It starts with an anticlockwise 
loop for the first peak (\emph{red portion}), then it traverses clockwise path for the second (\emph{blue}) and
it goes anticlockwise for the third peak (\emph{pink}). The clockwise loops represent soft lag whereas the 
anticlockwise loops indicate hard lag (\cite{KRM98}).     
This is possibly due to the fact that during the fast rise of the spike, more particles are 
accelerated quickly to higher energies by some acceleration process. During the decay part of the spike 
the cooling of the particles dominates and generates soft lag in the 
spectral evolution. This feature is very well reflected in the hardness--count rate loops shown in Figures 
\ref{hr1} -- \ref{hr4}. The clockwise loops arise
when the evolution of spectral slope is primarily governed by any cooling process which is faster
for higher energy particles (\cite{KRM98}). For GRB061121 anti-clockwise loops, signifying hard lag, appear when the 
acceleration time scale is nearly equal to the cooling time scale and particles are gradually accelerated
to higher energies (\cite{KRM98}). These behaviours of spectral evolution of GRBs strongly depend on the particle acceleration
process and the radiation emission mechanism occuring in the GRB outflow. With the present understanding of particle
acceleration process in turbulence, it is not completely understood what kind of particle acceleration and cooling 
process can really explain such properties along with the evolution of the exact spectral forms.
\end{itemize}
\begin{figure}
\begin{center}
\includegraphics[width=0.12\textheight,angle=-90]{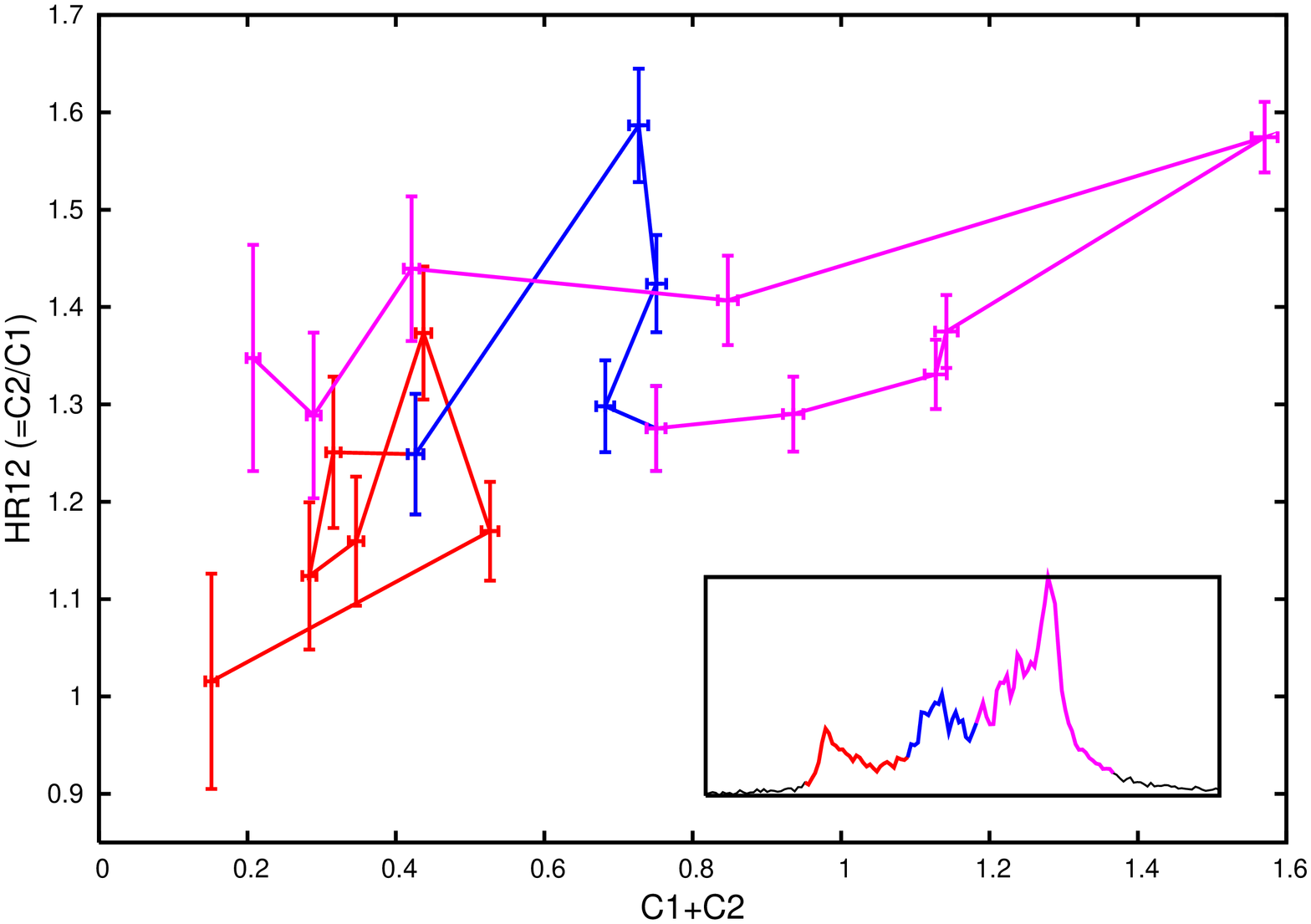}
\includegraphics[width=0.12\textheight,angle=-90]{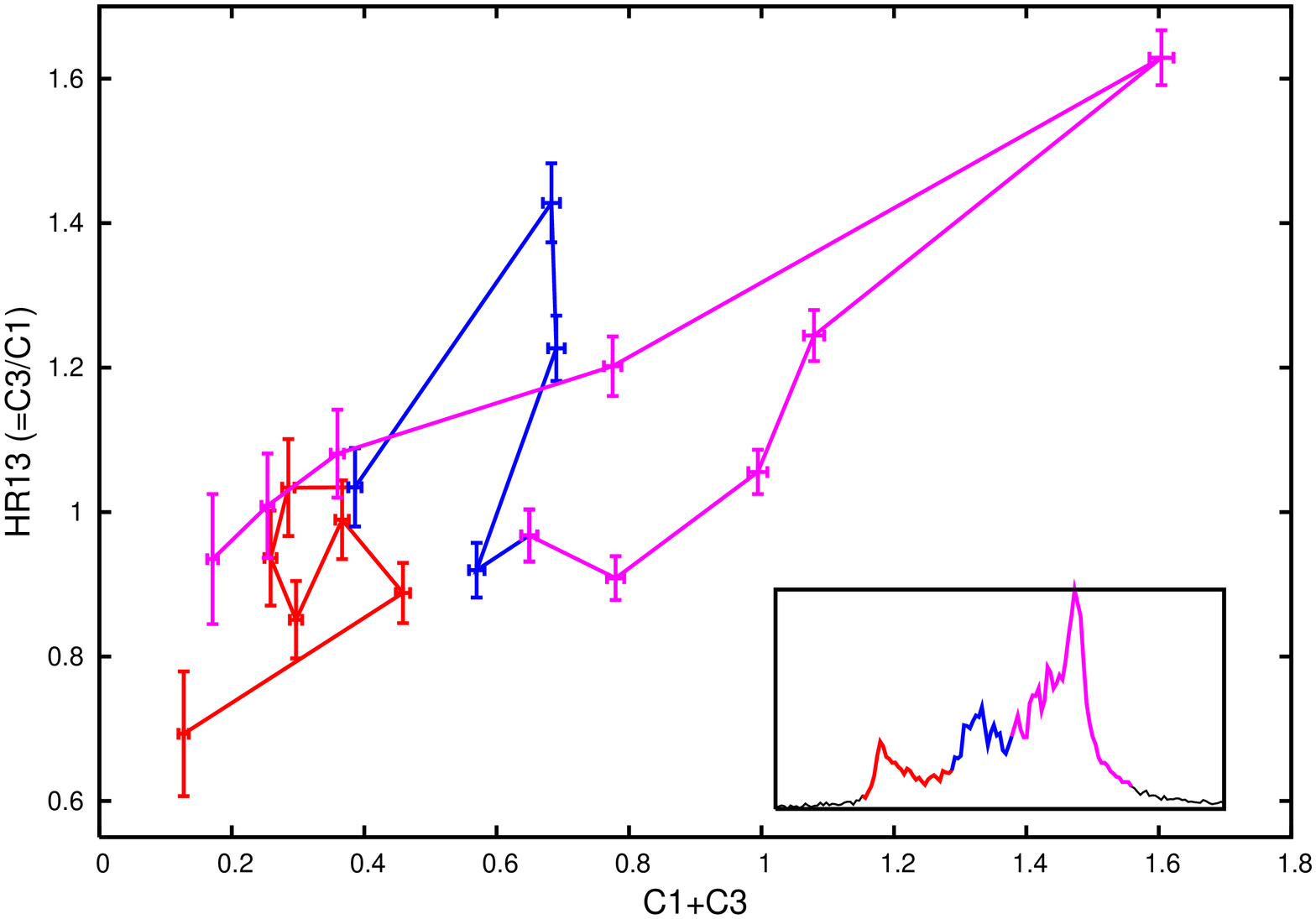}
\caption{GRB061121: same as Figure \ref{hr1}.} \label{hr3}
\end{center}
\end{figure}
\begin{figure}
\begin{center}
\includegraphics[width=0.12\textheight,angle=-90]{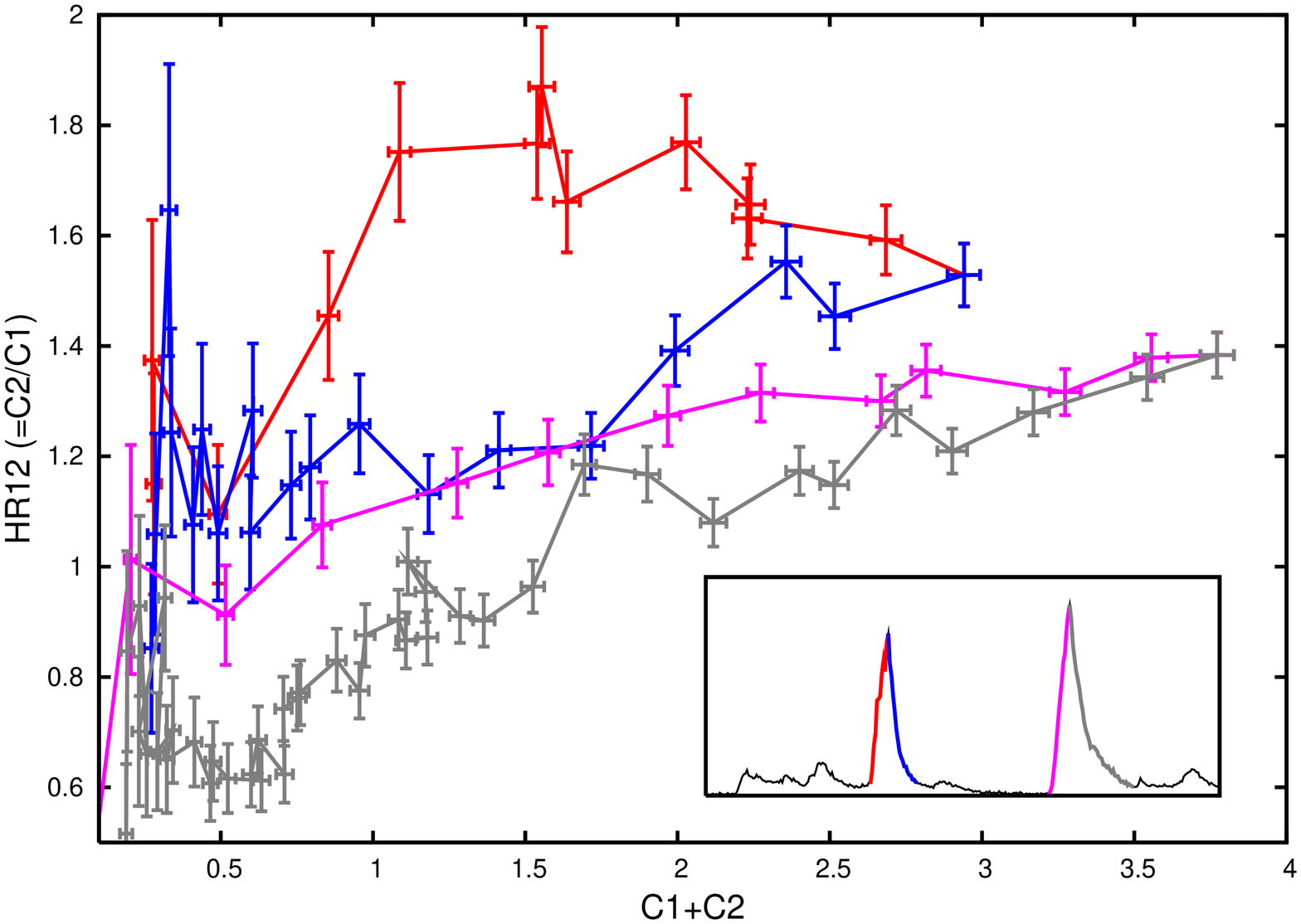}
\includegraphics[width=0.12\textheight,angle=-90]{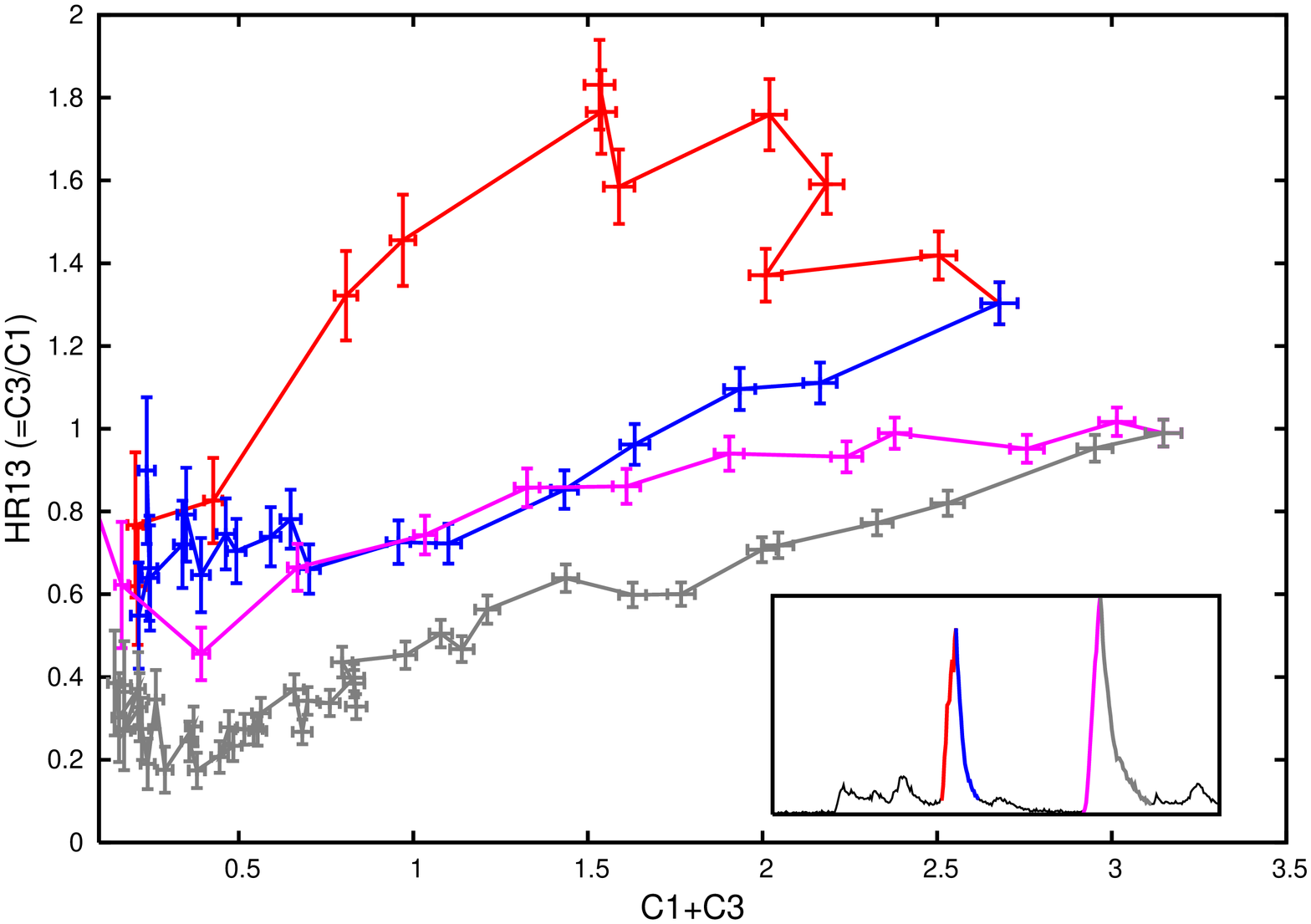}
\caption{GRB080411: same as Figure \ref{hr1}.} \label{hr4}
\end{center}
\end{figure} 

\section{Conclusion}
In this paper we studied the time-dependent probability density function obtained from the observed lightcurves of
four bright \emph{Swift} gamma-ray bursts. The probability density functions are fitted with Gaussian distribution
function. The parameters of the distributon function vary with time as the burst evolves. Considering the fact that
the short time-scale flux variations in GRBs are due to the random motion of radiation emitting turbulent eddies, as
proposed by \cite{NK09}, we argued that the Gaussian probability density function actually reflects the 
presence of turbulence in GRB outflow during the prompt emission.

\noindent We also studied the spectral properties of GRBs, such as the time evolution of spectral index, colour--colour
diagram and hysteresis loops to understand if they can be explained in the framework of turbulence. With the present 
understanding of particle acceleration in turbulence, it is found that the particle acceleration process in turbulent
recconection can partially explain the observed spectral indices in 15--150 keV energy range. The behaviour reflected
in colour--colour diagrams are consistant with spectral evolution seen for all GRBs. But deeper understanding of colour--
colour diagrams and hysteresis loops require detailed knowledge of different time scales, therefore a detailed study of
radiative processes in GRBs is essential and is beyond the scope of this paper.

\noindent Finally, we would like to remark that for steady sources like X-ray binaries (\emph{e.g.} Cyg X-1, (\cite{WBS10,Utt05})) and flaring
systems like solar flares (\cite{Zh07}), the probability density functions are found to follow log-normal distributions. For both types
of systems turbulence seems to a major role. In the light of such observations the main finding of this work, \emph{i.e.}
the Gaussian probability density functions for GRBs needs further study.

\section{Acknowledgement}
Authors acknowledge Sudhir Jain for fruitful discussion on the probability density functions, Abhas Mitra for suggesting
a relevent literature and the anonymous referee for very useful suggestions which enhanced the focus of the paper.

\bibliographystyle{mn2e}
\bibliography{ms-grb}

\end{document}